\documentclass[a4paper,11pt]{article}
\pdfoutput=1

\usepackage{jheppub1}
\usepackage[T1]{fontenc}
\usepackage{amssymb}
\usepackage{graphicx}
\usepackage{amsmath}
\usepackage{hyperref}
\usepackage{tensor}
\usepackage{epstopdf}
\usepackage{extarrows}
\newcommand{\td}{\text{d}}
\def \R {{^{(2)}R}}
\def \pD {\mathfrak{D}}

\def \hpD {\hat{\mathfrak{D}}}

\begin{document}

\title{Universal bounds on the size of a black hole}

\author{Run-Qiu Yang and H. L\"{u} }
\emailAdd{aqiu@tju.edu.cn}
\emailAdd{mrhonglu@gmail.com}
\affiliation{Center for Joint Quantum Studies and Department of Physics, School of Science, Tianjin University, Yaguan Road 135, Jinnan District, 300350 Tianjin, P.~R.~China}

\abstract{
For static black holes in Einstein gravity, if matter fields satisfy a few general conditions, we conjecture that three characteristic parameters about the spatial size of black holes, namely the outermost photon sphere  area $A_{\mathrm{ph,out}}$, the corresponding shadow area $A_{\mathrm{sh,out}}$ and the horizon area $A_{\mathcal{H}}$ satisfy a series of universal inequalities $9A_{\mathcal{H}}/4\leq A_{\mathrm{ph,out}}\leq A_{\mathrm{sh,out}}/3\leq 36\pi M^2$, where $M$ is the ADM mass. We present a complete proof in the spherically symmetric case and some pieces of evidence to support it in general static cases. We also discuss the properties of the photon spheres in general static spacetimes and show that, similar to horizon, photon spheres are also conformal invariant structures of the spacetimes.
}

%
%
%
\maketitle
%
%

\section{Introduction}
Black holes are fundamental objects in Einstein's general relativity. The spatial size of a black hole is usually characterized by its horizon; however, the horizon cannot be directly observed in classical theories either locally or from asymptotic infinity. A few of recent arguments (e.g.~see Refs.~\cite{Almheiri:2012rt,Avery:2012tf}) suggest that quantum effects may render the horizon locally observable, but this topic remains controversial. There is another special surface named ``photon sphere'' where gravity is also so strong that photons are forced to travel in orbits~\cite{Bardeen1972,Hod_2013,Cardoso_2009}. Differing from the horizon, some photons can escape from the photon sphere, making it observable. The photon sphere plays a key role for gravitational lensing~\cite{Virbhadra:1999nm,Stefanov:2010xz} or ringdown of waves around a black hole~\cite{Cardoso:2016rao}. It is also related to the characteristic (quasinormal) resonances
of black-hole spacetimes~\cite{Cardoso_2009,Hod:2009td,Yang_2012,Hod:2012bw}. For a Schwarzschild black hole of mass $M$, the radius of photon sphere is $3M$. The outmost photon sphere is unstable and can cast a ``shadow'' for an observer at the asymptotic infinity. Recently, the first picture of a black hole shadow was taken~\cite{Akiyama:2019cqa}, which gave us a direct impression of the appearance of the black hole size and shape.

Owing to the significance in astrophysical observations, it is important to study the photon spheres and their shadows. Although the classical properties of the horizon have been well studied, the photon spheres and shadows are still lack of extensive investigations. In a spherically symmetric black hole of mass $M$, Hod proved that for Einstein gravity coupled to matter satisfying the weak energy condition and negative trace energy condition, the innermost photon sphere radius $r_{\mathrm{ph,in}}$ and total mass $M$ satisfy~\cite{Hod:2017xkz}
\begin{equation}\label{hodineq}
  r_{\mathrm{ph,in}}\leq 3M.
\end{equation}
By using the same energy condition, Ref.~\cite{Cvetic:2016bxi} proved an relationship between innermost photon sphere and its shadow radius: $r_{\mathrm{sh,in}}\geq\sqrt{3}r_{\mathrm{ph,in}}$.
A lower bound $r_{\mathrm{ph,in}}\geq 2M$ was conjectured also by Hod~\cite{Hod:2012nk} but counterexample was found by Ref.~\cite{Cvetic:2016bxi}.

For the observational purpose, it is more relevant to consider the outermost photon sphere. The proof of Hod's does not apply to the outermost one when there are multiple photon spheres, which do exist in black holes satisfying the dominant energy condition~\cite{liu2019quasitopological}. Recently, a series of universal inequalities about outermost photon sphere was proposed~\cite{Lu:2019zxb,Feng:2019zzn}
\begin{equation}
\label{unverineq}
  3r_{+}/2\leq r_{\mathrm{ph,out}}\leq r_{\mathrm{sh,out}}/\sqrt{3}\leq 3M.
\end{equation}
Here $r_+$ is the radius of the horizon. Refs.~\cite{Lu:2019zxb,Feng:2019zzn} verify it in many different black holes. Its generalization to higher dimensions were discussed in~\cite{ma2019bounds}.

In this paper, we will first prove the inequalities \eqref{unverineq} for spherically symmetric and static black holes in Einstein gravity, for matter fields satisfying a few simple requirements.  We then consider
more general static configurations and define the corresponding ``photon sphere'' and ``outermost'' photon sphere.  We conjecture that the area of outermost photon sphere $A_{\mathrm{ph,out}}$, the corresponding shadow area $A_{\mathrm{sh,out}}$ and horizon area $A_{\mathcal{H}}$ (if exists), also satisfy a series of universal inequalities, sandwiched within the Penrose inequality:
\begin{equation}\label{unverineq2}
  9A_{\mathcal{H}}/4\leq A_{\mathrm{ph,out}}\leq A_{\mathrm{sh,out}}/3\leq 36\pi M^2.
\end{equation}
Although we do not have the full proof of Eq.~\eqref{unverineq2} yet, we will give some pieces of evidence to support it. We will also show that, similar to the horizon, photon spheres in static spacetimes are also conformal invariant structures.




\section{Spherically symmetric case}
We first present the full proof of \eqref{unverineq} for spherically symmetric metrics in $(3+1)$ dimensions, which read
\begin{equation}\label{RNmetric}
  \td s^2=-f(r)e^{-\chi(r)}\td t^2+\frac{\td r^2}{f(r)}+r^2\td\Omega_2^2\,.
\end{equation}
Here $f(r_+)=0$, $f(r)$ is positive when $r>r_+$. Einstein's equation reduces to the following three equations,
\begin{eqnarray}\label{eqforfchi}
&&f'=-8\pi r\rho+\frac{1-f}{r},\qquad\chi'=-\frac{8\pi r}{f}(\rho+p_r)\,,\\
\label{constrT0}
&&p_r'=\frac1{2fr}[{\cal N} (\rho+p_r)+2fT-8fp_r]\,,
\end{eqnarray}
where $\mathcal{N}:=3f-1-8\pi r^2p_r$, and $p_r$, $\rho$, $T$ are radial pressure, energy density and the trace of stress tensor respectively. We require $r^3p_r(r)\rightarrow0$ and $r^3\rho(r)\rightarrow0$ when $r\rightarrow\infty$. Photon spheres are determined by $U'=0$, where
\begin{equation}\label{defU1}
  U(r):=f(r)e^{-\chi(r)}/r^2\,,\quad \hbox{and}\quad U'=-\mathcal{N} e^{-\chi}/r^3\,.
\end{equation}
It is clear that $U$ must have an extremum. Multiple and odd numbers of extrema can also arise and the radii of  photon spheres satisfy $\mathcal{N}=0$\cite{Hod:2017xkz}. Furthermore, we must have $\mathcal{N}>0$ and $U'<0$ if $r>r_{\mathrm{ph,out}}$.

\subsection{Proof of upper bounds}
Here we show that if both the weak and strong energy conditions are satisfied for $r\ge r_{\mathrm{ph,out}}$, we have
\begin{equation}\label{ineqpart1}
  r_{\mathrm{ph,out}}\leq r_{\mathrm{sh,out}}/\sqrt{3}\leq 3M\,.
\end{equation}
The radius of shadow $r_{\mathrm{sh,out}}$ is related to the outermost photon sphere by $r_{\mathrm{sh,out}}=1/\sqrt{U(r_{\mathrm{ph,out}})}$~\cite{Cvetic:2016bxi}.

We introduce an auxiliary function $\mathcal{W}(r):=e^{-\chi(r)}[1+8\pi r^2p_r(r)]$, which has following relations
\begin{equation}\label{bounfr}
  U(r_{\mathrm{ph,out}})=\mathcal{W}(r_{\mathrm{ph,out}})/(3r_{\mathrm{ph,out}}^2),~~(r^3U)'=\mathcal{W}\,.
\end{equation}
The key is to show that $\mathcal{W}(r)\leq1$ when $r\geq r_{\mathrm{ph,out}}$. The Einstein's equation and null energy condition imply $\chi\geq0$. The weak energy condition tells us $\rho\geq0$. Using Eq.~\eqref{eqforfchi}, we see $\max f=1-8\pi r^2 \rho|_{f'=0}\leq1$. We now split the interval $[r_{\mathrm{ph,out}},\infty)$ into two groups: $\{I_1^+, I_2^+,\cdots\}$ where $p_r\geq0$ and $\{I_1^-,I_2^-,\cdots\}$ where $p_r\leq0$. If $r\in I_n^-$ (here $n=1,2,\cdots$), we see $\mathcal{W}(r)\leq1$. If $r\in I_n^+$, i.e., $p_r\geq0$, we find that the derivative of $\mathcal{W}(r)$ is
\begin{equation}\label{dUdrs1}
  \mathcal{W}'(r)=4\pi r e^{-\chi}[(\rho+p_r)(1+8\pi r^2p_r+f)/f+4p_T]\,.
\end{equation}
Here $p_T:={T^{\theta}}_{\theta}={T^{\phi}}_{\phi}$ is the transverse pressure.
As $p_r\geq0$ and $f\leq1$, we see that $\mathcal{W}'\geq4\pi r e^{-\chi}[(\rho+p_r)(1+f)/f+4p_T]\geq 8\pi r e^{-\chi}(\rho+p_r+2p_T)$. Thus the strong energy condition ensures that $\mathcal{W}(r)$ is non-decreasing in the interval $I_n^+$. The maximal value of $\mathcal{W}(r)$ at every interval $I_n^+$ is theqrefore at the endpoint, where $p_r=0$. Thus, we also have $\mathcal{W}(r)\leq1$ at the interval $I_n^+$. We can now immediately see $r_{\mathrm{sh,out}}\geq \sqrt{3}r_{\mathrm{ph,out}}$.


To prove the shadow upper bound, we introduce $\widetilde U=(1-2M/r)/r^2$ for the Schwarzschild black hole of the same mass. It is clear that $(r^3\widetilde{{U}})'=1\geq\mathcal{W}= (r^3{U})'$.
At $r\rightarrow\infty$, the asymptotical flatness requires matter fields to decay at least in following ways
\begin{equation}\label{matterseq}
  \rho\sim\mathcal{O}(1/r^{3+\delta_1}),~~p_r\sim\mathcal{O}(1/r^{3+\delta_2}),~~\delta_1>0,~~\delta_2>0\,.
\end{equation}
Then we find
$$\chi=\mathcal{O}(r^{-(1+\min\{\delta_1,\delta_2\})}),~~f=1-2M/r+\mathcal{O}(r^{-(1+\delta_1)})\,,$$
which implies
\begin{equation}\label{matterseq2}
  U=\frac{fe^{-\chi}}{r^2}=r^{-2}-\frac{2M}{r^3}+\mathcal{O}(r^{-(3+\min\{\delta_1,\delta_2\})})\,.
\end{equation}
Then we have $[r^3\widetilde{{U}}(r)-r^3{U}(r)]|_{r\rightarrow\infty}=0$.
Thus, we have $\widetilde{{U}}(r)\leq {U}(r)$ when $r\geq r_{\mathrm{ph,out}}$. As  the maximum of $\widetilde{U}$ in the interval $[r_{\mathrm{ph,out}},\infty)$ is $1/(27M^2)$, we see the upper bound
\begin{equation}\label{RRs2q2}
1/(27 M^2)=\max\widetilde{U}\leq\max U=1/r_{\mathrm{sh,out}}^2.
\end{equation}
In addition,  we have the rigidity: for all spherically symmetric static spacetimes of mass $M$, $r_{\mathrm{ph,out}}=3M$ or $r_{\mathrm{sh,out}}=3\sqrt{3}M$ arises if and only if the exterior of photon sphere is the Schwarzschild.

If we only focus on the photon sphere, the requirement can be much relaxed and we only need the null energy condition outside the outermost photon sphere.
Our main tool is a new mass function
\begin{equation}\label{deHawm1}
  M(r):=\mathfrak{m}(r,\rho)+\frac{4\pi}3 r^3p_r(r)=\frac{r}3 (1-\mathcal{N}/2)\,.
\end{equation}
Here $\mathfrak{m}(r,\rho)$ is the Hawking-Geroch mass~\cite{doi:10.1063/1.1664615,Geroch1973}, given by
\begin{equation}\label{hawkingm1}
  \mathfrak{m}(r,\rho):=\frac{r}2[1-f(r)]=r_+/2+4\pi\int_{r_+}^rx^2\rho(x)\td x\,.
\end{equation}
Applying Eqs.~\eqref{eqforfchi} and \eqref{constrT0} we find an identity
\begin{equation}\label{dMdr}
  M'(r)=\frac{8\pi r^2}3(\rho+p_T)+\frac{2\pi r^2}{3f}(\rho+p_r)\mathcal{N}.
\end{equation}
For $r\geq r_{\mathrm{ph,out}}$, $\mathcal{N}\geq0$, the null energy condition ensures $M'\geq0$. This shows $M(r_{\mathrm{ph,out}})\leq M(\infty)=M$.  At the photon sphere $\mathcal{N}=0$, we have $M(r_{\mathrm{ph,out}})=r_{\mathrm{ph,out}}/3$.
Thus, we obtain $r_{\mathrm{ph,out}}\leq 3M$. Compared to Hod's proof~\eqref{hodineq}, our condition is much weaker but the conclusion is stronger (as $r_{\mathrm{ph,out}}\geq r_{\mathrm{ph,in}}$). This result also implies a new positive mass theorem and entropy bound for static spherically symmetric black holes: if the null energy condition outside the black hole is satisfied, the mass $M$ must be positive and the horizon radius must be smaller than $3M$ (cannot saturate this bound). In fact, the condition can be even weaker: There exists a photon sphere outside which the null energy condition is satisfied. This is remarkable since even in the spherical case, the previous proofs often require the weak energy condition.

Stronger inequalities may be obtained for some particular matters. For example, if the matters contain the standard Maxwell field with charge $Q$ and all the other matters satisfy null energy condition, then Eq.~\eqref{dMdr} implies $M'\geq2Q^2/(3r^2)$ and so $M(r_{\mathrm{ph,out}})\leq M-2Q^2/(3r_{\mathrm{ph,out}})$, which leads to a tighter bound in the charged case $r_{\mathrm{ph,out}}\leq \frac{3M}2(1+\sqrt{1-8Q^2/(9M^2)})$. Thus, in all spherically symmetric black holes of same mass and charge, Reissner-Norstr\"om (RN) black hole has largest photon sphere.

\subsection{Proof of lower bound}
It follows from Eq.~\eqref{hawkingm1} that $\mathfrak{m}(r,\rho)\ge r_+/2$ when $\rho\geq0$, then we see from Eq.~\eqref{deHawm1} that the lower bound $3r_+/2\leq r_{\mathrm{ph,out}}$ holds if weak energy condition holds and $p_r\ge 0$ at $r=r_{\mathrm{ph}}$. This is generally satisfied by astronomical black holes.

In theory, many important solutions such as the RN black holes have negative $p_{r}$. In these cases, the weak energy condition alone is not enough to ensure $3r_+/2\leq r_{\mathrm{ph,out}}$.
As an example, we consider $\chi=0$ and
\begin{equation}
f(r)=1-\frac{r_+}{r}-\frac{6\rho_0r_+}{\sqrt{e}r}+\rho_0e^{-r/(2r_+)}(2+4r_+/r),
\end{equation}
for which
$\rho(r)=-p_r(r)=\frac{\rho_0}{8\pi rr_+}e^{-r/(2r_+)}$
with $\rho_0>0$, satisfying the weak energy condition.  After specifying $\rho_0=1$,   we find $r_{\mathrm{ph,out}}/r_+\approx1.417<3/2$.
However, we find that the lower bound holds if $p_r$ and $\rho$ satisfy an additional condition: there is a function $\Xi(r)$ such that,
    \begin{equation}\label{addpr1}
      \forall r>r_+,~~~[r^2\Xi(r)]'\geq0~~\mathrm{and}~-\rho\leq\Xi(r)\leq p_r(r)\,.
    \end{equation}
This requirement is weak in the sense that we only need the existence of such function $\Xi(r)$. This condition admits the negative pressure. For example, in Einstein-Maxwell theory, $\rho(r)=-p(r)\propto Q^2/r^4$ and one we can take $\Xi(r)=p(r)$. The proof is as follows.

$r^2\Xi(r)$ is a non-decreasing function outside the horizon and the null energy condition implies $\rho\geq-\Xi$ and $p_r\geq \Xi$ and so $M(r)\geq \mathfrak{m}(r,-\Xi)+\frac{4\pi r^3}3\Xi$.
On the other hand, we find
\begin{equation}\label{eqwu1}
  \mathfrak{m}(r,-\Xi)=r_+/2-4\pi\int_{r_+}^rx^2\Xi(x)\td x\geq{r_+}/2-4\pi r^2\Xi(r)(r-r_+)\,,
\end{equation}
which gives us
\begin{equation}\label{rhmrplusf2}
  M(r)+8\pi r^3\Xi(r)/3\geq [1+8\pi r^2\Xi(r)]\, r_+/2\,.
\end{equation}
%
Substituting $r=r_{\mathrm{ph,out}}$ and $M(r_{\mathrm{ph,out}})=r_{\mathrm{ph,out}}/3$ into the above, we obtain
\begin{equation}\label{eqfph1}
  \left(r_{\mathrm{ph,out}} - 3{r_+}/2\right) [1+8\pi r_{\mathrm{ph,out}}^2\Xi(r_{\mathrm{ph,out}})]\geq 0\,.
\end{equation}
As $r_+$ is the outermost horizon, we have $f'(r_+)\geq0$. Eq.~\eqref{eqforfchi} implies $1-8\pi r^2_+\rho(r_+)\geq0$ and theqrefore
\begin{equation}\label{reqxis2}
  1+8\pi r_+^2\Xi(r_+)\geq0\Rightarrow1+8\pi r^2\Xi(r)\geq0\,
\end{equation}
if $r>r_+$. We thus prove the lower bound. This proof also applies to the stronger statement $r_{\mathrm{ph,in}}\geq 3r_+/2$.


\section{General static cases}
\subsection{Generalization of photon spheres}
In this part we consider the general static spacetimes,
where it is more instructive to study the ``photon sphere'' in the spacetime rather than only to focus on its spatial projection. In static spacetime, there is a timelike Killing vector $\xi^\mu=(\partial/\partial t)^\mu$ outside the horizon and $t$ is the time coordinate. The metric of spacetime has following 3+1 decomposition,
\begin{equation}\label{decomsigam1a}
  \td s^2=-\phi^2\td t^2+h_{ab}\td x^a\td x^b\,.
\end{equation}
Here $h_{ab}$ is the metric of equal-$t$ slice $\Sigma_t$ and $\phi^2=\xi^\mu\xi_\mu$. Motivated by Ref.~\cite{Yoshino_2017}, we call a connected timelike co-dimensional 1 surface $\Gamma=\{t\}\times\mathcal{S}$ to be marginal transversely-trapping surface (MTTS), if $\mathcal{S}$ is topological 2-sphere and any null geodesic that starts tangentially on $\Gamma$ will keep laying on $\Gamma$. Let $n^\mu$ be its outward unit normal covector, $T^\mu$ be the tangent vector of a null geodesic. We see $T^\mu n_\mu|_\Gamma=0$ and so $T^\mu\nabla_\mu(n_\nu T^\nu)=0$. We thus have the condition for an MTTS~\cite{Claudel_2001}:
\begin{equation}\label{condmtts}
  \forall~\mathrm{null~tangent~vector }T^\mu, ~~K_{\mu\nu}T^\mu T^\nu=0\,.
\end{equation}
Here $K_{\mu\nu}$ is extrinsic curvature of the MTTS. 
Then an equal-$t$ cross-section $\mathcal{S}$ is a photon sphere.

\begin{figure}
  \centering
  \includegraphics[width=0.4\textwidth]{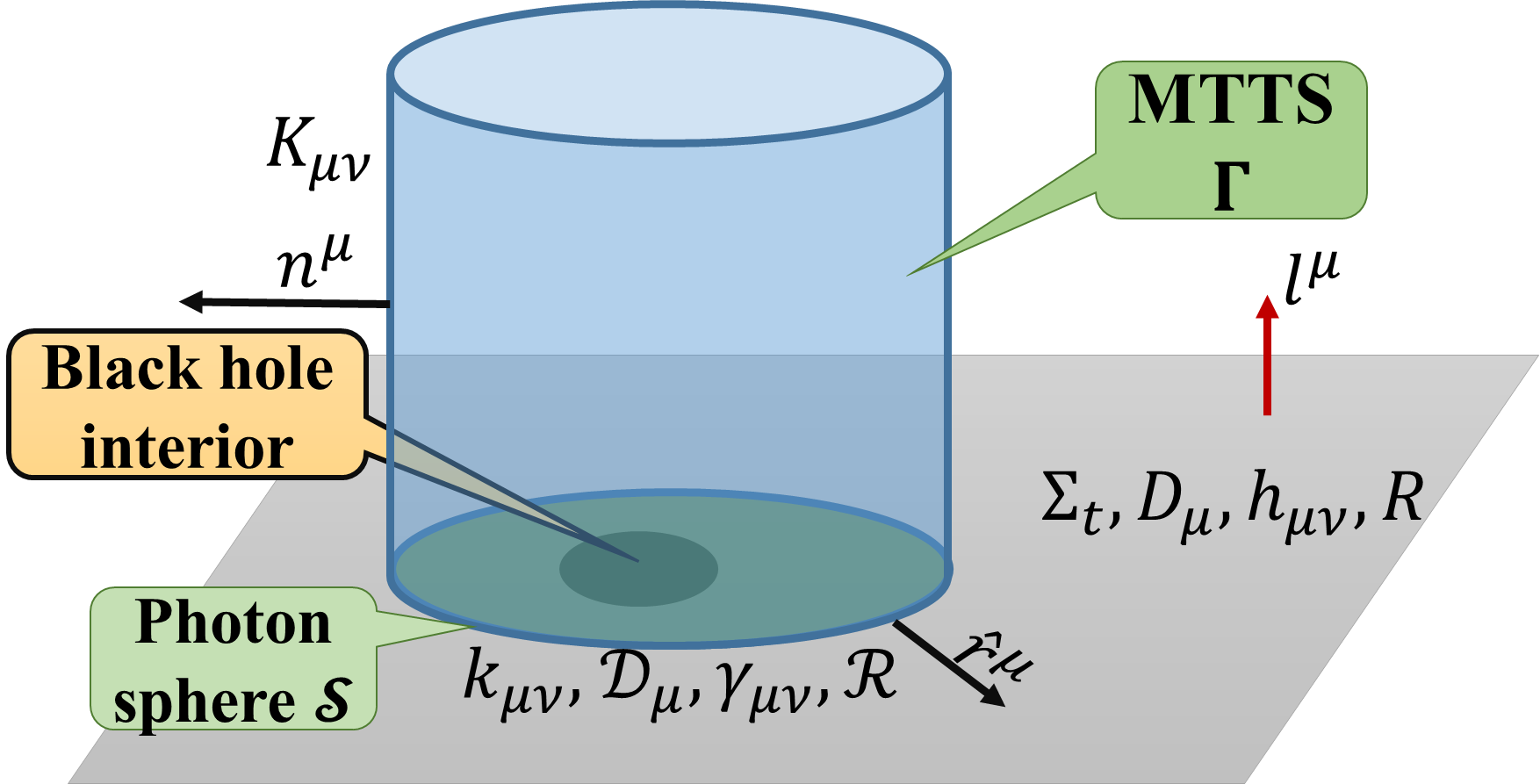}
  \caption{Notations in the MTTS and photon sphere. For convenience, we immerse all the vectors/tensors into the 3+1 dimensional spacetime and so their indexes become spacetime indexes. For example, $h_{\mu\nu}={e^a}_\mu{e^b}_\nu h_{ab}$, where ${e^a}_\mu$ is the pull-black map from spacetime to the equal-$t$ surface. } \label{FigMTTS}
\end{figure}

We can give a more explicit expression to find a photon sphere in the static spacetime. Assume that $\mathfrak{R}$ is the scalar curvature of $\mathcal{S}$, $\Sigma_t$ is an equal-$t$ slice and $R$ is its scalar curvature, $k_{\mu\nu}$ is the extrinsic curvature of $\mathcal{S}$ embedded in $\Sigma_t$  and its trace is $k$, $l^\mu$ is the unit normal vector of $\Sigma_t$ and $(\partial/\partial t)^\mu=\phi l^\mu$, $\hat{r}^\mu$ is the outward unit normal vector of $\mathcal{S}$ embedded in $\Sigma_t$. $(\gamma_{\mu\nu}, \mathcal{D}_\mu) $ and $(h_{\mu\nu}, D_\mu)$ are the induced metrics and covariant derivatives of the photon sphere $\mathcal{S}$ and static slice $\Sigma_t$, respectively. See Fig.~\ref{FigMTTS} for schematic explanations on these notations.
In static cases, one can find that $n^\mu|_{\mathcal{S}}=\hat{r}^\mu|_{\mathcal{S}}$ and so we have the decomposition $K_{\mu\nu}=-l_{\mu}l_{\nu}\phi^{-1}\hat{r}^\tau D_\tau\phi+k_{\mu\nu}$.
Assuming that $s^\mu$ is an arbitrary unit tangent vector field of $\mathcal{S}$, then $T^\mu=l^{\mu}+s^\mu$ is a null vector tangent to MTTS. The requirement~\eqref{condmtts} implies $\phi^{-1}\hat{r}^\mu D_\mu\phi-k_{\mu\nu}s^\mu s^\nu=0$. As the result we find
\begin{equation}\label{eqphks1}
  k_{\mu\nu}=\gamma_{\mu\nu}\phi^{-1}\hat{r}^\mu D_\mu\phi\,.
\end{equation}
%
Using the decomposition of the Einstein's tensor
\begin{equation}\label{prEinst}
  \mathfrak{R}=-16\pi p_r+2\phi^{-1}\mathcal{D}^2\phi+2k\phi^{-1}\hat{r}^\mu D_\mu\phi+k^2-k_{\mu\nu}k^{\mu\nu}\,,
\end{equation}
where $p_r:=T_{\mu\nu}\hat{r}^\mu\hat{r}^\nu$ is the pressure on $\mathcal{S}$, we can obtain
\begin{equation}\label{decomp1d}
  \left. 3k^2/4-8\pi p_r+\phi^{-1}\mathcal{D}^2\phi-\mathfrak{R}/2\right|_{\mathcal{S}}=0\,.
\end{equation}
It reduces to $\mathcal{N}=0$ in the spherical case.

Similar to the horizon, the photon sphere is also a conformal invariant structure. This can be understood by the fact that the null geodesics are conformal invariant, or that Eq.~\eqref{eqphks1} is invariant under the conformal transformation $\{h_{\mu\nu}\rightarrow\tilde{h}_{\mu\nu}=\Omega^2 h_{\mu\nu}, ~\phi\rightarrow\tilde{\phi}=\Omega\phi\}$. Particularly, if we choose $\Omega=\phi^{-1}$, then $\tilde{h}_{\mu\nu}$ is just the ``optical metric''~\cite{Cvetic:2016bxi} and the trace of extrinsic curvature is $\tilde{k}|_{\mathcal{S}}=0$. Thus, photon sphere $\mathcal{S}$ is a minimal surface in the ``optical metric''. For a surface $S$, assume that $\td S$ is the surface element induced by original metric $h_{\mu\nu}$. Then the surface element under the optical metric reads $\phi^{-2}\td S$. Thus, photon sphere is a critical surface which locally minimizes following functional
\begin{equation}\label{defBeq1}
  P:=\int_S\phi^{-2}\td S\,.
\end{equation}
It should be pointed out that we generalized the photon sphere concept directly from the well-defined spherical case. However, whether such {\it thin-shell} photon sphere exists in general remains to be further investigated. It may be necessary for us to introduce some ``weak photon spheres'' by relaxing requirement~\eqref{condmtts} while keeping most of the essential properties. See e.g.~Ref.~\cite{cao2019quasilocal} for an example.  Nevertheless, we shall proceed with the assumption.

\subsection{Outermost photon sphere and conjectures about its size}
\begin{figure}
  \centering
  \includegraphics[width=0.29\textwidth]{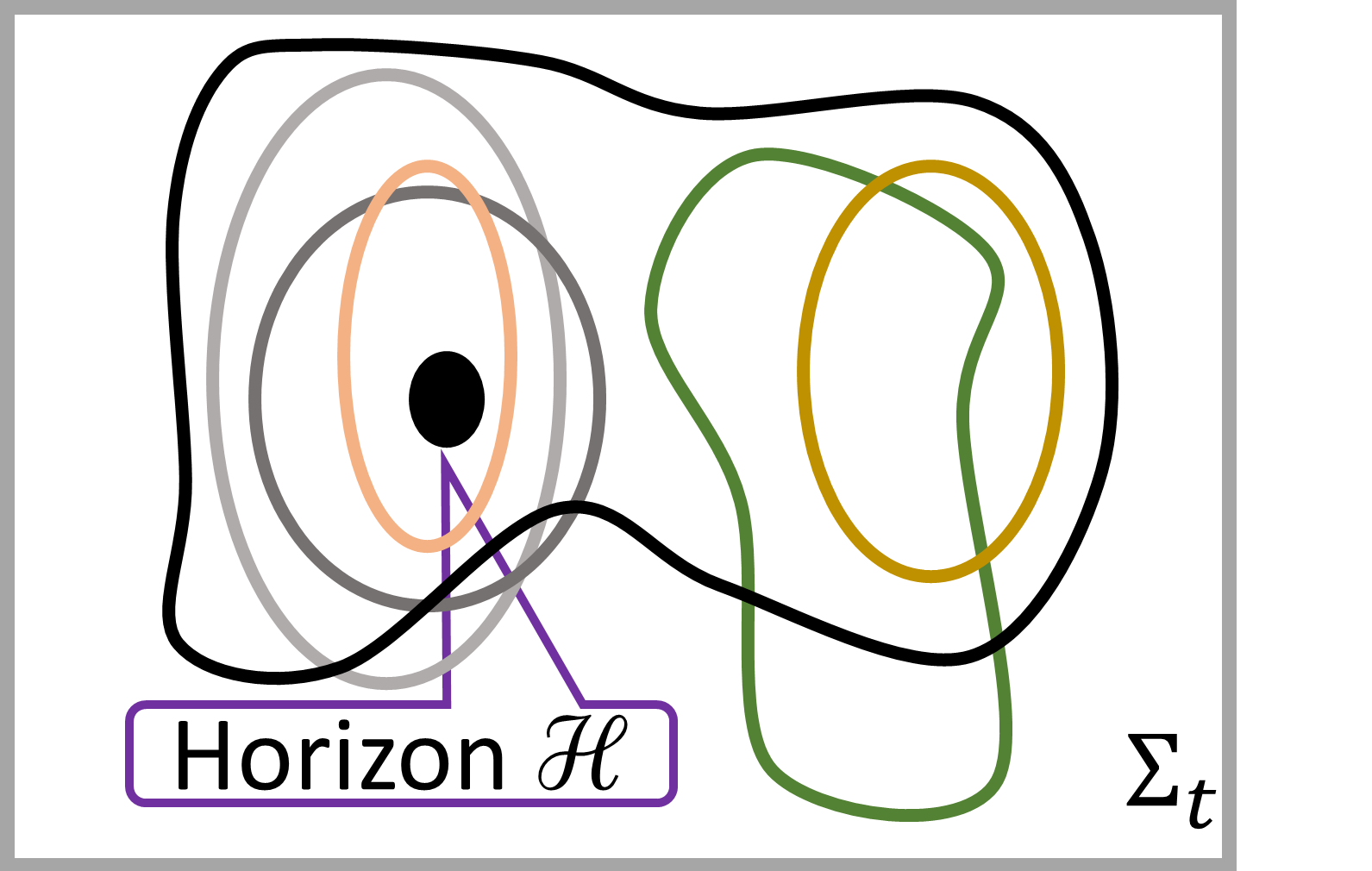}
  \includegraphics[width=0.29\textwidth]{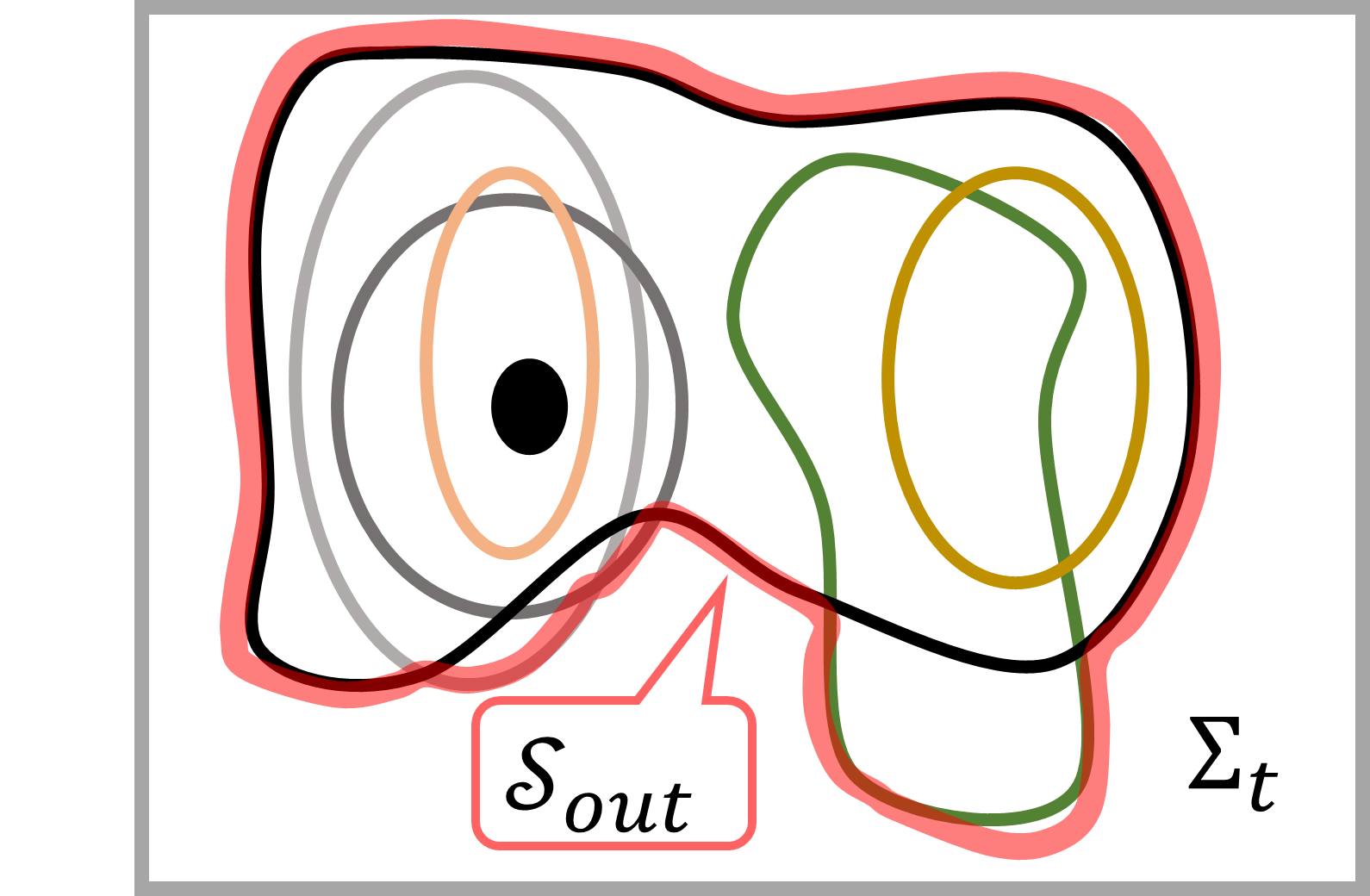}
  \includegraphics[width=0.39\textwidth]{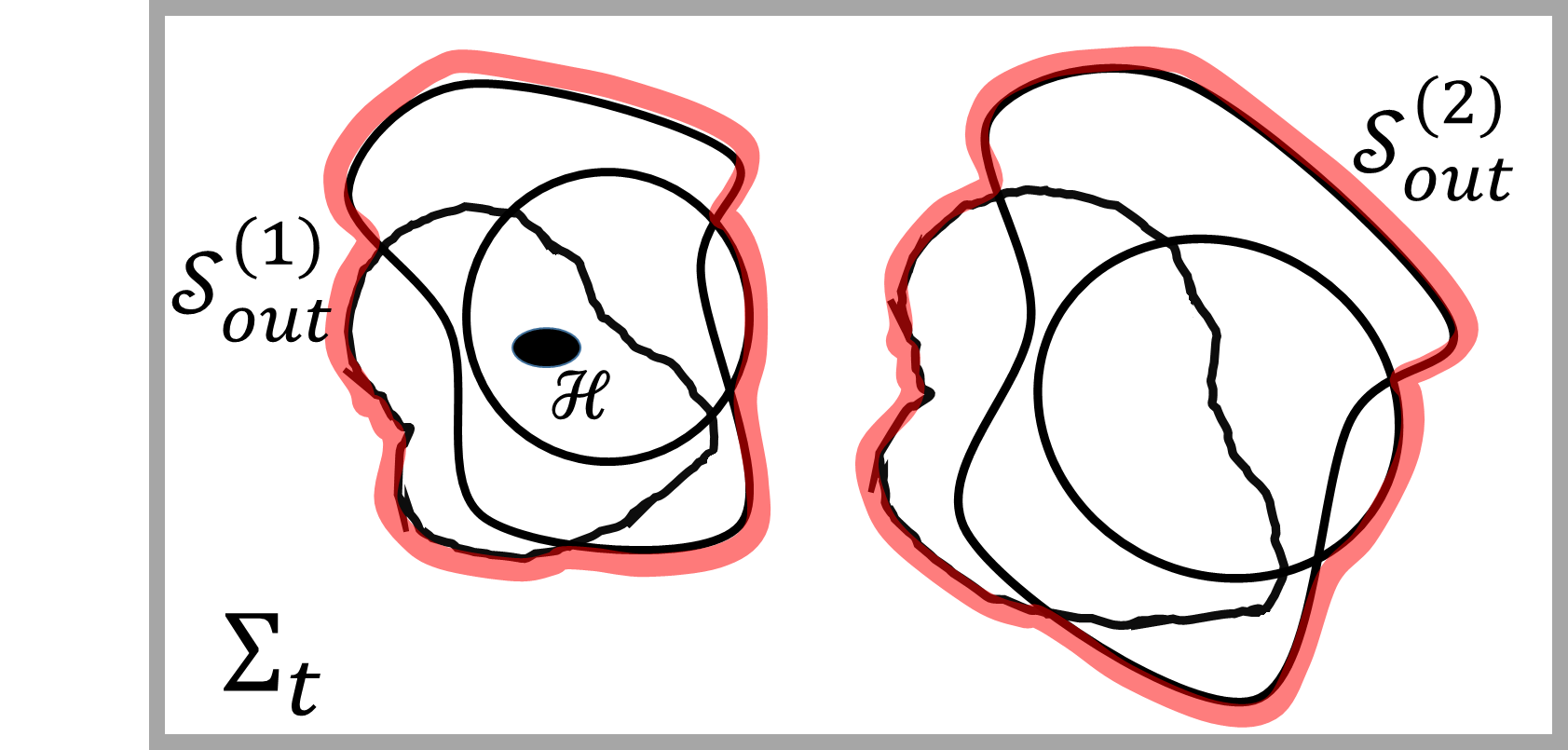}
  \caption{Left: the circles stand for different photon spheres, which may intersect with each others. Middle: red circle is the ``outermost photon sphere'', which is just the enveloping surface of outermost segments of photon spheres. Right: the outermost photon sphere contains two connected branches $\mathcal{S}_{\mathrm{out}}^{(1)}$ and $\mathcal{S}_{\mathrm{out}}^{(2)}$, and $\mathcal{S}_{\mathrm{out}}=\mathcal{S}_{\mathrm{out}}^{(1)}\cup\mathcal{S}_{\mathrm{out}}^{(2)}$.  } \label{FigMTTS2}
\end{figure}

For general static spacetimes, the photon spheres may intersect with each others and have many inequivalent homology classes. See the left panel of Fig.~\ref{FigMTTS2} for example. The meaning of the ``outermost'' photon sphere needs to be clarified. We propose a proper definition about the ``outermost'' should satisfy the following four requirements: (1) it satisfies Eq.~\eqref{eqphks1} piecewise and no tangentially null geodesic can escape outside; (2) it is closed; (3) no any part of photon spheres is outside it; and (4) $\forall$ topological 2-sphere $X$ outside the ``outermost'' photon sphere,  we have
\begin{equation}\label{decomp2}
  \frac34k^2-8\pi p_r+\phi^{-1}\mathcal{D}^2\phi-\mathfrak{R}/2|_X>0\,,
\end{equation}
where $k, \mathcal{D}^2$ and $\mathfrak{R}$ are the trace of extrinsic curvature, Laplace operator and scalar curvature of $X$ respectively, and $p_r$ is the pressure normal to $X$. In the spherical case, Eq.~\eqref{decomp2} recovers the condition $\mathcal{N}>0$. Based on these considerations, we define the outermost photon sphere $\mathcal{S}_{\mathrm{out}}$ as the enveloping surface of outermost segments of all photon spheres, illustrated in the middle panel of Fig.~\ref{FigMTTS2}.
The $\mathcal{S}_{\mathrm{out}}$ may be disconnected and contain many connected branches $\mathcal{S}_{\mathrm{out}}^{(i)}$, i.e., $\mathcal{S}_{\mathrm{out}}=\bigcup_{i}\mathcal{S}_{\mathrm{out}}^{(i)}$. See the right panel of Fig.~\ref{FigMTTS2} for example.

We denote the area of $\mathcal{S}_{\mathrm{out}}^{(i)}$ to be $A_{\mathrm{ph,out},i}$, i.e.
\begin{equation}\label{ashadow1}
  A_{\mathrm{ph,out},i}=\int_{\mathcal{S}_{\mathrm{out}}^{(i)}}\td S\,.
\end{equation}
The $\mathcal{S}_{\mathrm{out}}^{(i)}$ will cast a shadow at the observer's sky. In the spherical case, the shadow is a disk, of which the radius is independent of the angle of view. In general cases, the shadow may have complicated shapes and depend on the angle of view. It is more convenient to study the apparent area of photon sphere measured at infinity, which is given by following integration,
\begin{equation}\label{ashadow1}
  A_{\mathrm{sh,out},i}=\int_{\mathcal{S}_{\mathrm{out}}^{(i)}}\phi^{-2}\td S\,.
\end{equation}
Here $\td S$ is the surface element induced by original metric $h_{\mu\nu}$.  We can use $A_{\mathrm{sh,out},i}$ to characterize the size of shadow. In the spherically symmetric case $r_{\mathrm{sh,out}}=\sqrt{A_{\mathrm{sh,out},i}/(4\pi)}$.
Assume $A_{H,i}$ to be the area of horizon inside $\mathcal{S}_{\mathrm{out}}^{(i)}$. The inequalities in~\eqref{ineqpart1} have a naturally generalization:
\begin{equation}\label{goalb}
  9A_{H,i}/4\leq A_{\mathrm{ph,out},i}\leq A_{\mathrm{sh,out},i}/3\leq 36\pi M^2\,,
\end{equation}
We also conjecture a global version which involves the union of all the connected branches:
\begin{equation}\label{goalb2}
  9A_{\mathcal{H}}/4\leq A_{\mathrm{ph,out}}\leq A_{\mathrm{sh,out}}/3\leq 36\pi M^2\,,
\end{equation}
Here $A_{\mathcal{H}}=\sum_i A_{H,i}$ and the same for others.

Although we do not have a complete proof beyond the spherical case, we can already prove some parts now in special situations. For example, $9A_{\mathcal{H}}/4\leq36\pi M^2$ is simply the Penrose inequality and has been proven by several different methods~\cite{Mars:2009cj}. In following we will offer the proofs on  $9A_{\mathcal{H}}/4\leq A_{\mathrm{sh,out}}/3$ and $A_{\mathrm{ph,out}}\leq A_{\mathrm{sh,out}}/3\leq 36\pi M^2$ in the special case without the spherical symmetry.

\section{Proofs without spherically symmetry}
As the first step to consider the general case, we assume that the outmost horizon is \textit{connected}, and always assume the \textit{weak and strong} energy conditions. We also assume that outmost photon sphere is \textit{smooth} at beginning. Finally, we will give argument to show that the smoothness of outmost photon sphere can be relaxed into \textit{piecewise smoothness}. We will convert the inequalities into the problems of finding maximum/minimum, which can be solved by variational method. In general, to check if the variational problem will give us minimum or maximum, we may need to compute the second order variation. This in general will be complicated. However, if the on-shell value of variational problem has explicit expression, we have simpler way.

Assume a functional is $\mathcal{F}:\mathcal{X}\mapsto \mathbb{R}$, where $\mathcal{X}$ is the space spanned by all arguments of $\mathcal{F}$.  Mathematically, if $\mathcal{F}$ satisfies a few of very general restrictions~\footnote{In mathematical jargon, the necessary conditions are (1) functional $\mathcal{F}$ is weak coercivity in $\mathcal{X}$ and (2) $\mathcal{F}$ is weak upper/lower semicontinuity, e.g. see Ref.~\cite{Bernard}. We always assume that those requirements on target functional are satisfied in physics.} and bounded from above/below, then $\mathcal{F}$ must has maximum/minimum~\cite{Bernard} and the maximum/minimum of $\mathcal{F}$ must be given by the maximal/minimal on-shell value of corresponding variational problem. Thus, to find the maximum/minium of $\mathcal{F}$, we only need to find all on-shell values and pick out the maximal/minimal one. In the case that the on-shell value of variational problem has explicit expression, this is easy to perform.

\subsection{Bondi-Scahs formalism}\label{BSF1}
We will use Bondi-Scahs formalism~\cite{Sachs1962,Madler:2016xju} in this section, which offers us simple way to study behaviors of metric in non-spherical case. Bondi-Scahs formalism foliates the spacetime by a series of null surfaces which are labeled by $u=$constant and the general metric has following form
\begin{equation}\label{BSmetric}
  \td s^2=-\frac{V}{r}e^{2\beta}\td u^2-2e^{2\beta}\td u\td r+r^2h_{AB}(\td x^A-U^A\td u)(\td x^B-U^B\td u)\,.
\end{equation}
As the spacetime is asymptotically flat, we then fix the boundary conditions as follows
\begin{equation}\label{boundcs}
  \beta|_{r\rightarrow\infty}=0,\quad\quad \left.\frac{V}{r}\right|_{r\rightarrow\infty}=1,\quad\quad h_{AB}|_{r\rightarrow\infty}\td x^A\td x^B=\td\Omega^2\,,
\end{equation}
where $\td\Omega^2$ is the metric of unit sphere.
The photon spheres are geometrical structures of a static black hole, so they are independent of the choices of coordinates. To give readers clear visible pictures, we used static 3+1 decomposition. To study the inequalities proposed in this paper, the Bondi-Scahs coordinates gauge is more convenient.  

\begin{figure}
  \centering
  \includegraphics[width=0.45\textwidth]{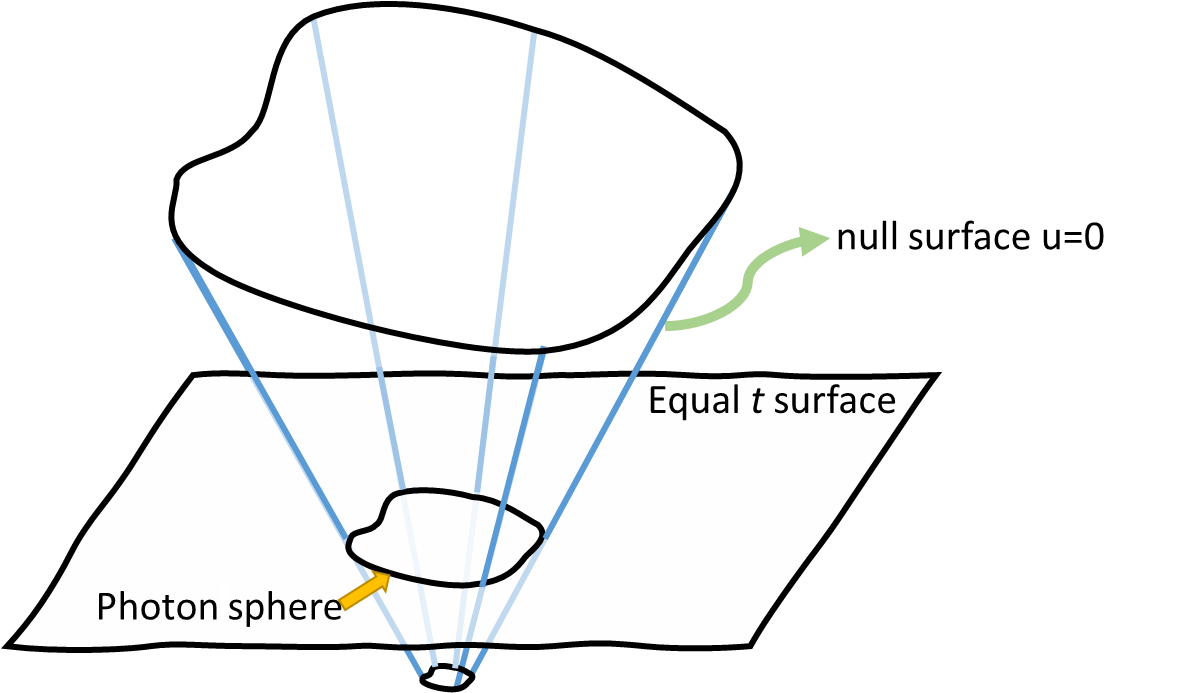}
  \includegraphics[width=0.45\textwidth]{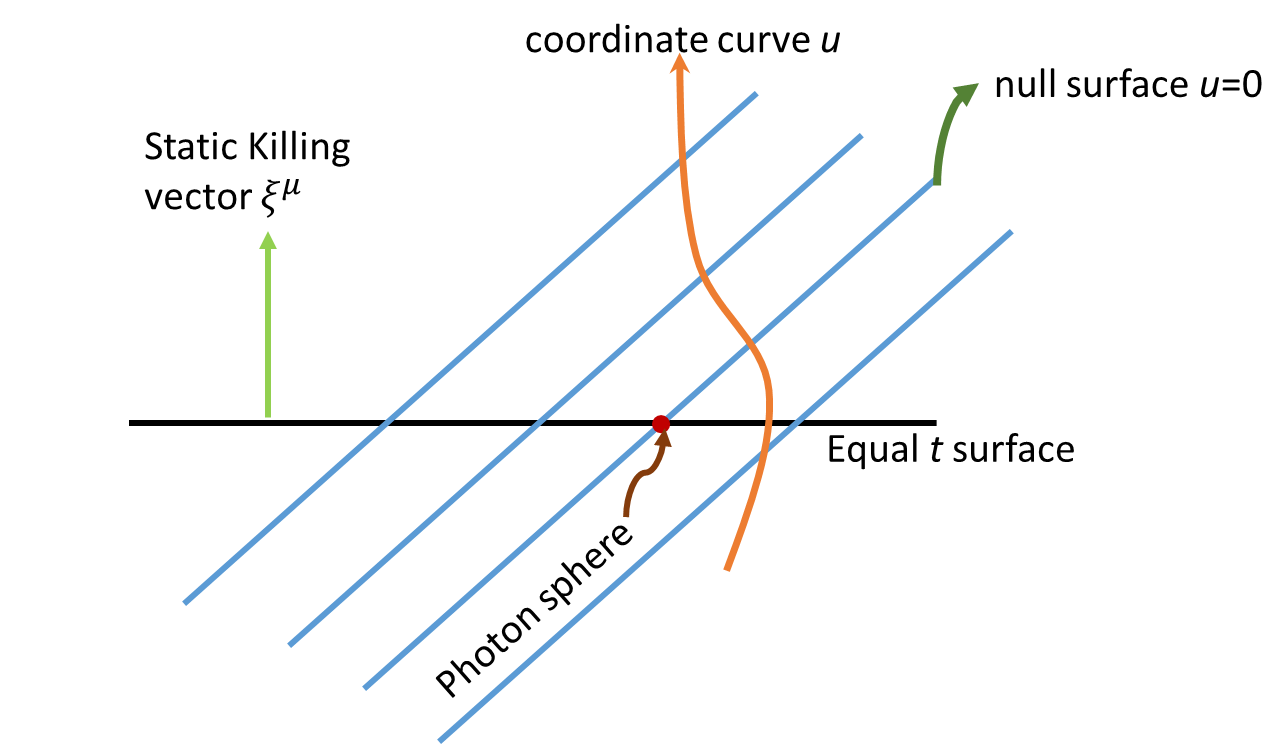}
  \caption{Left: the outgoing null rays of photon sphere form the null surface $u=0$. A photon sphere $\mathcal{S}$ locates at an equal-$t$ spacelike surface as well as locates at a null sheet. On the null sheet, the photon sphere then is defined by $r=r_\mathcal{S}(x^A)$. Right:  the null foliation of spacetime. Other null sheets are obtained by using translation generated by static Killing vector $\xi^\mu$. The $u$-coordinate is free and we can choose arbitrary timelike curve which passes through these null sheets. } \label{Fignull1}
\end{figure}

In  Bondi-Scahs gauge, we have three gauge freedoms. Firstly, we have freedom to choose where the null hypersurface $u=0$ locates. One convenient choice is that we use outgoing light rays of a photon sphere $\mathcal{S}$ to define the null hypersurface $u=0$. See Fig.~\ref{Fignull1}.
The photon sphere $\mathcal{S}$ then locates at equal-$t$ surface as well as locates at $u=0$ null sheet. Thus, $\mathcal{S}$ is also defined by $u=0$ and $r=r_\mathcal{S}(x^A)$. After we obtain the first null sheet $u=0$, all the other null sheets can be obtained by using the translation generated by static Killing vector $\xi^\mu$. Every null sheet is labeled by a constant $u$. In fact, besides photon sphere, for any closed surface $S$ in the equal-$t$ submanifold and outside horizon, we can always the immerse one closed surface $S$ into its outgoing light rays. Using this initial null sheet and the translation generated by $\xi^\mu$, we can always obtain a local null foliation and $S$ lays on one null sheet. 

Secondly, we have freedom to choose $u$-coordinates. The tangent vector of $u$-coordinate is $(\partial/\partial u)^\mu$. In Bondi-Scahs gauge, the normal covector of null sheets $\propto(\td u)^\mu$ but normal vector is \emph{not} proportional to $(\partial/\partial u)^\mu$. In fact, the normal vector of equal-$u$ surface satisfies
\begin{equation}\label{normaleq1}
  g^{\mu\nu}(\td u)_\nu\propto (\partial/\partial r)^\mu\,,
\end{equation}
which lays on the equal-$u$ null sheets. This defines the coordinate $r$ and we find
$$g_{rr}=g_{\mu\nu}(\partial/\partial r)^\mu(\partial/\partial r)^\nu=0\,,$$
which matches with the metric~\eqref{BSmetric}.  The $u$-coordinate is arbitrary timelike curve and passes through those null sheets, e.g. see Fig.~\ref{Fignull1}. We can choose $(\partial/\partial u)^\mu=\xi^\mu$, i.e. the static Killing vector field. Then we have
\begin{equation}\label{phiVbeta}
  \phi^2=\xi^\mu\xi_\mu=g_{uu}=Ve^{2\beta}/r-r^2h_{AB}U^AU^B\,.
\end{equation}
Here $\phi$ is same as \eqref{decomsigam1a}.

The third gauge freedom comes from the fact that the relationship~\eqref{normaleq1} does not fix the choice of coordinate $r$ uniquely. We can choose $\tilde{r}$ to replace $r$, where $\tilde{r}$ satisfies $(\partial/\partial r)^\mu=\psi(\partial/\partial r)^\mu$ with arbitrary nonzero scalar field $\psi$. To eliminate this freedom, we can impose gauge condition $\partial_rh=0$. By this gauge choice, we see
\begin{equation}\label{horizonA0}
  \sqrt{h}\td^2x=\sqrt{h}\td^2x|_{r\rightarrow\infty}=\td\Omega\,.
\end{equation}
and $\td\Omega$ is a surface element of unite sphere. The differential gauge $\partial_rh=0$ leaves a freedom in choosing the initial condition of $r$. To fix this freedom, we can set the horizon radius $r_h$ to be constant, which leads to a simple formula for the area of horizon
\begin{equation}\label{horizonA}
  A_{\mathcal{H}}=4\pi r_h^2\,.
\end{equation}
Alternatively, we can choose a photon sphere $\mathcal{S}$ to be constant $r=r_{\mathrm{ph}}$, then the area of $\mathcal{S}_{\mathrm{ph}}$ has simple expression,
\begin{equation}\label{horizonA}
  A_{\mathrm{ph}}=4\pi r_{\mathrm{ph}}^2\,.
\end{equation}
In general, we \textit{cannot} require that both horizon and photon sphere have constant $r$.

The coordinates $\{x^A\}$ lay on the null sheets. We consider the static case, so we can choose the transverse direction $x^A$ to be orthogonal to $\xi^\mu$, which means $U^A=0$. Then metric becomes
\begin{equation}\label{BSmetric2}
  \td s^2=-\frac{V}{r}e^{2\beta}\td u^2-2e^{2\beta}\td u\td r+r^2h_{AB}\td x^A\td x^B\,.
\end{equation}
In appendix~\ref{staticBS} we offer more detailed argument on Eq.~\eqref{BSmetric2}. Strictly speaking, in some spacetimes such coordinates may only cover a finite neighborhood. Here we assume that it can cover the whole region outside horizon $\mathcal{H}$.

In above coordinates gauge, the Einstein's equation shows~\cite{Madler:2016xju}
\begin{equation}\label{eqforbeta}
  \partial_r\beta=\frac{r}{16}h^{AC}h^{BD}(\partial_rh_{AB})(\partial_rh_{CD})+2\pi r T_{rr}
\end{equation}
and
\begin{equation}\label{eqforV}
  \partial_r(e^{-2\beta}V)=\frac{\R}2-\pD^2\beta-(\pD\beta)^2-8\pi r^2\rho-\frac{Vre^{-2\beta}}{8}h^{AC}h^{BD}(\partial_rh_{AB})(\partial_rh_{CD})\,.
\end{equation}
Here $\R$ and $\pD_A$ are the scalar curvature and covariant derivative operator of $h_{AB}$. $T_{rr}:=T_{\mu\nu}(\partial/\partial r)^\mu(\partial/\partial r)^\nu$ is the null-null component of energy momentum tensor $T_{\mu\nu}$, $\rho:=T_{\mu\nu}\xi^\mu \xi^\nu/(\xi^\mu\xi_\mu)$. The weak energy condition then insures
\begin{equation}\label{weakeq1a}
  T_{rr}\geq0,~~\rho\geq0\,.
\end{equation}
Then Eq.~\eqref{eqforbeta} implies
\begin{equation}\label{boundbeta}
  \beta\leq0\,.
\end{equation}
In following we will define some target functionals on some surfaces $S$, which will depend on the spacetime geometry and choices of $S$. We always assume the Einstein's equation, so the spacetime geometry is determined by distribution of matters. As the result, the functionals will depend on distribution of matters and choices of $S$. It is necessary to know how many degrees of freedoms the target functionals have. The energy momentum tensor has 10 components, 4 of which are constrained by Bianch identity (or the conserved law $\nabla^\mu T_{\mu\nu}=0$). The requirement $U^A=0$ also supplies 2 constraints according to Einstein's equation. Thus, we have 4 bulk degrees of freedoms. Ten components of energy momentum tensor combining with the choice of surface $S$ offers us 11 degrees of freedoms at the surface $S$. $U^A|_S=0$ and $h|_S=h|_{r\rightarrow\infty}$ offer us 3 constraints on the surface $S$. Thus, we we have 8 degrees of freedoms at a surface $S$. Theqrefor, we can choose at most 4 bulk variables and at most 8 surface variables as the independent variables when we use variational method. It will simplify the variational problem if we choose the independent variables suitably.

\subsection{Proof of $9A_{\mathcal{H}}/4\leq A_{\mathrm{sh,out}}/3$}
In this subsection, we will use variational method to prove: in all black holes of which outmost horizon areas are $A_{\mathcal{H}}$, if photon sphere exists, then the minimum value of $A_{\mathrm{sh,out}}$ is $27A_{\mathcal{H}}/4$. To do that, we introduce an auxiliary functional
\begin{equation}\label{defLS0}
  \mathcal{B}:=A_{S}^{-2}\int_S\phi^2\td S\,.
\end{equation}
Here $S$ is arbitrary surface which encloses outmost horizon $\mathcal{H}$ and lays on the equal-$t$ surface, $A_{S}$ is the area of $S$. The Cauchy-Schwartz inequality shows that
$$\left(\int_S\phi^2\td S\right)\int_S\phi^{-2}\td S\geq\left(\int_S\phi\phi^{-1}\td S\right)^2=A_{S}^2\,$$
and so
\begin{equation}\label{cseqab1}
  \int_S\phi^{-2}\td S\geq1/\mathcal{B}\,.
\end{equation}
Then for outmost photon sphere $\mathcal{S}_{\mathrm{sh,out}}$, we have
\begin{equation}\label{relAb1}
  A_{\mathrm{sh,out}}=\int_{S_{\mathrm{sh,out}}}\phi^{-2}\td S\geq\frac1{\mathcal{B}|_{S=\mathcal{S}_{\mathrm{sh,out}}}}\geq\frac1{\max\mathcal{B}}\,.
\end{equation}
In appendix~\ref{uppbd} we show that, if we fix the horizon area and strong energy condition is satisfied, then $\mathcal{B}$ is bounded from above. Thus, functional $\mathcal{B}$ has maximum. To find the this maximum, we need to find all extreme values and pick up the maximal one. For every surface $S$ lay on the equal-$t$ submanifold, we can immerse it into the null sheet $u=0$ and parameterize it by $\{u=0,r=r_S(x^A)\}$ in the Bondi-Scahs coordinates gauge. In the appendix~\ref{extrms1}, we have shown that: if a surface $S$ makes $\mathcal{B}$ extreme among all surfaces which lay on equal-$t$ surface, it will also makes $\mathcal{B}$ extreme among all surfaces which lay on $u=0$ null sheet.  Then we have following relationship
\begin{eqnarray}\label{maxbs1}
  &~&(\max \mathcal{B}\mathrm{~on~}\Sigma_t)=(\mathrm{maxiaml~extreme~value~of~}\mathcal{B}\mathrm{~on~}\Sigma_t)\nonumber\\
  &=&(\mathrm{one~of~extreme~values~of~}\mathcal{B}\mathrm{~on~null~sheet~}u=0)\\
  &\leq&(\mathrm{maximal~extreme~value~of~}\mathcal{B}\mathrm{~on~null~sheet~}u=0)\nonumber\,.
\end{eqnarray}
Assume that $\mathcal{B}_{m}$ is the the maximal critical value of $\mathcal{B}$ on the null sheet $u=0$,\footnote{It does not mean that $\mathcal{B}_m$ is the maximal value of $\mathcal{B}$ on the null sheet $u=0$. Though $\mathcal{B}$ on the equal-$t$ hypersurface is shown to be bounded from above, it may have no upper bound on the null sheet $u=0$. }, then we see that max$~\mathcal{B}$ in Eq.~\eqref{relAb1} satisfies $\max\mathcal{B}\leq\mathcal{B}_{m}$. In the null sheet $u=0$, if we can show $\mathcal{B}_{m}=4/(27A_{\mathcal{H}})$, then  $A_{\mathrm{sh,out}}\geq27A_{\mathcal{H}}/4$ is a corollary. This is what we will do in this subsection.

In the null sheet $=0$, the value of $ \mathcal{B}$ depends on the function $r_S(x^A)$ as well as the metric and matter distributions. In following we will use variational method to find the maximum of $ \mathcal{B}$ on the null sheet $u=0$.  As we have analyzed at the end of Sec.~\ref{BSF1}, we can choose at most 4 bulk variables and at most 8 surface variables as the independent variables when we use variational method. We will first do not specify what are independent variables but only use Einstein's equation to rewrite $ \mathcal{B}$ into a different form (see Eq.~\eqref{defy1}) and then specify what are independent variables, by which we can easy to perform variation and find the on-shell values.

In appendix~\ref{exampf1} we use a concrete function as an example to explain the main idea of following steps. Though what we will do in following is more complicated than the example of appendix~\ref{exampf1}, the mathematical essence has no difference. Referring to the metric~\eqref{BSmetric2}, our target functional reads
\begin{equation}\label{defLS}
  \mathcal{B}=A_{S}^{-2}\int_S r^{-1}Ve^{2\beta}\td S\,.
\end{equation}
We take the gauge that the horizon has constant radius $r=r_h$. Then we find
\begin{eqnarray}\label{valueV1}
  Ve^{-2\beta}|_{r=r_0}&=&\int_{r_h}^{r_0}[\R/2-\pD^2\beta-(\pD\beta)^2\\
&-&\left.8\pi r^2\rho-\frac{Vre^{-2\beta}}{8}h^{AC}h^{BD}(\partial_rh_{AB})(\partial_rh_{CD})\right]\td r\nonumber\,.
 \end{eqnarray}
In general $Ve^{2\beta}$ will depend on $r_0$ and $x^A$ both. The induced metric on $S$ reads $\td s_S^2=r_S^2h_{AB}\td x^A\td x^B$, so we find $\td S=r_S^2\td\Omega$. Eqs.~\eqref{valueV1} and Eq.~\eqref{defLS} then show
\begin{eqnarray}\label{valueV1g}
  \mathcal{B}&=&\frac1{A_S^2}\int\td\Omega r_Se^{4\beta}\int_{r_h}^{r_S}\left[\R/2-\pD^2\beta-(\pD\beta)^2\right.\\
  &-&\left.8\pi r^2\rho-\frac{Vre^{-2\beta}}{8}h^{AC}h^{BD}(\partial_rh_{AB})(\partial_rh_{CD})\right]\nonumber\,.
\end{eqnarray}
and area of $S$ is
\begin{equation}\label{areaAs1}
  A_{S}=A_{S}[r_S]:=\int r_S^2\td\Omega\,.
\end{equation}
We define,
\begin{equation}\label{defineb2}
  \cos\Phi_S(x^A):=e^{4\beta}|_S\,,
\end{equation}
and
\begin{equation}\label{defineM1}
  N_1(r,x^A)^2:=8\pi r^2\rho,~~{N_2}(r,x^A)^2:=\frac{Vre^{-2\beta}}{8}h^{AC}h^{BD}(\partial_rh_{AB})(\partial_rh_{CD})
\end{equation}
which take the constraints $\beta\leq0, \rho\geq0$ and $V\geq0$ into account.
Then we find
\begin{equation}\label{valueV1f}
  \mathcal{B}=\frac1{A_S^2}\int\td\Omega r_S\cos\Phi_S\int_{r_h}^{r_S(x^A)}\left[\R/2-\pD^2\beta-(\pD\beta)^2-N_1^2-N_2^2\right]\td r\,.
\end{equation}
It needs to note that $r_S(x^A)$ in general may not be constant. To relax the dependence of upper limit $r_S$ in the second integration of Eq.~\eqref{valueV1f}, we can introduce step-function $\Theta(x)$ so that
\begin{eqnarray}\label{generalSA3b}
  \mathcal{B}&=&\frac{1}{A_{S}[r_S]^2}\int\td\Omega r_S\cos\Phi_S\int_{r_h}^{\infty}\td r\Theta(r_S-r)\\
  &\times&\left[\R/2-\pD^2\beta-(\pD\beta)^2-N_2^2-N_1^2\right]\nonumber\\
  &=&\frac{1}{A_{S}[r_S]^2}\int_{r_h}^{\infty}\td r\int\td\Omega \cos\Phi_S r_S\Theta(r_S-r)\nonumber\\
  &\times&\left[\R/2-\pD^2\beta-(\pD\beta)^2-N_2^2-N_1^2\right]\nonumber\,.
\end{eqnarray}
As $h_{AB}$ is metric of 2-dimensional space, we can always find a coordinate transformation $x^A\rightarrow y^A$ such that
\begin{equation}\label{cofrg1}
  h_{AB}\td x^A\td x^B=e^{2\Psi(r,y^A)}\gamma_{AB}\td y^A\td y^B\,,
\end{equation}
where $\gamma_{AB}$ is the standard metric of unit sphere. By using this conformal metric, we find that $\mathcal{B}$ becomes
\begin{eqnarray}\label{defy1}
  \mathcal{B}&=&\frac{1}{A_{S}[r_S]^2}\int_{r_h}^{\infty}\td r\int r_S\cos\Phi_S\Theta(r_S-r)[1-\hpD^2\Psi\nonumber\\
  &-&\hpD^2\beta-(\hpD\beta)^2-e^{-2\Psi}N_1^2-e^{-2\Psi}N_2^2]\sqrt{\gamma}\td^2y\,.
\end{eqnarray}
Here $\hpD_A$ is the covariant derivative operator corresponding to $\gamma_{AB}$.

From Eq.~\eqref{defLS0} to \eqref{defy1}, we just made identical deformations according to Einstein's equation. In Eq.~\eqref{defy1}, we find that $\{\beta,r_S,\Phi_S,N_1, {N_2}, \Psi\}$ has 4 independent bulk variables and only 2 independent surface variables, so we can treat all them as independent variables. These are independent arguments of functional $\mathcal{B}$.

We solve the variational problem by following steps. The variation with respective to $\Psi$ is easy computed from Eq.~\eqref{valueV1f} and we obtain
\begin{equation}\label{eqforPsi0}
  \sin\Phi_S\int_{r_h}^{r_S}\left[\R/2-\pD^2\beta-(\pD\beta)^2-N_1^2-N_2^2\right]\td r=0\,.
\end{equation}
Using Eq.~\eqref{valueV1} and definitions of $N_1$ and ${N_2}$, we find that Eq.~\eqref{eqforPsi0} reduces
\begin{equation}\label{eqforPsi}
  V e^{-2\beta}|_{r=r_S}\sin\Phi_S=0\,.
\end{equation}
This equation shows that $\Phi_S=0$. The variations with respective to ${N_2}$ and $N_1$ read
\begin{equation}\label{eqformn1}
  e^{-2\Psi}r_S\Theta(r_S-r)N_1=e^{-2\Psi}r_S\Theta(r_S-r){N_2}=0\,
\end{equation}
and so ${N_2}=N_1=0$ when $r_h<r<r_S$. Then by using Eq.~\eqref{defy1}, we find that variation with respective to $\Psi$ gives us
\begin{equation}\label{rqforhab}
  \hpD^2[\Theta(r_S-r)r_S]=0,~~r_h<r<r_S\,.
\end{equation}
This equation shows that $\Theta(r_S-r)r_S$ must be independent of $y^A$. This is because
\begin{equation}\label{harmeq1}
  \forall f(y^A),~~\int f\hpD^2f\td\Omega=\int(\partial f)^2\td\Omega\,.
\end{equation}
Then we see $\hpD^2[\Theta(r_S-r)r_S]=0\Rightarrow\partial_A[\Theta(r_S-r)r_S]=0$. Thus Eq.~\eqref{rqforhab} shows $r_S$ to be constant. Then the variation with respective to $\beta$ shows
\begin{equation}\label{eqforbeta0}
  \hpD^2\beta=0\,,
\end{equation}
which shows that $\beta=\beta(r)$.  Thus, the variation with respective to $r_S$ shows
\begin{equation}\label{eqforLs1}
  4r_SA_{S}\mathcal{B}=\int_{r_h}^{\infty}[\Theta(r_S-r)+r_S\delta(r_S-r)]\td r\,,
\end{equation}
where
\begin{equation}\label{eqforLs1b}
  A_S=4\pi r_S^2,~~\mathcal{B}=\frac{r_S-r_h}{4\pi r_S^3}
\end{equation}
Eq.~\eqref{eqforLs1} shows $r_S=3r_h/2$. Finally, the on-shell value of $\mathcal{B}$ on the null sheet $u=0$
\begin{equation}\label{onshellL1}
  \mathcal{B}|_{\mathrm{on-shell}}=1/(27\pi r_h^2)=\frac4{27A_{\mathcal{H}}}\,.
\end{equation}
The value of $\Psi$ cannot be determined by variational method, so the solutions of variational problem are not unique. However, the on-shell values of all solutions are same. As Eq.~\eqref{onshellL1} is the only on-shell value of the variational problem, we can conclude that $4/(27A_{\mathcal{H}})$ is the maximal extreme value of $\mathcal{B}$ on the null sheet $u=0$. This proves $9A_{\mathcal{H}}/4\leq A_{\mathrm{sh,out}}/3$.

\subsection{Proof of $A_{\mathrm{ph,out}}\leq A_{\mathrm{sh,out}}/3$}\label{subseca2}
In this subsection, we do not fix the horizon area but fix the functional $\mathcal{B}|_{S=\mathcal{S}_{\mathrm{ph,out}}}=\mathcal{B}_0$. In appendix~\ref{uppbd}, we have shown $A_S<1/\mathcal{B}$ for any surface outside horizon and laying on equal-$t$ submanifold. If we fix $\mathcal{B}|_{S=\mathcal{S}_{\mathrm{ph,out}}}=\mathcal{B}_0$, the areas of photon spheres are bounded from above, so we can use variational method to find the maximum of $A_{\mathrm{ph,out}}$. We will prove
\begin{equation}\label{maxAbs1}
  \max A_{\mathrm{ph,out}}=\frac1{3\mathcal{B}_0}\,,
\end{equation}
If this is true, we then have
\begin{equation}\label{relb0s0}
  A_{\mathrm{ph,out}}\leq \frac1{3\mathcal{B}_0}=\frac1{3\mathcal{B}|_{S=\mathcal{S}_{\mathrm{ph,out}}}}\leq\frac13\int_{\mathcal{S}_{\mathrm{out}}}\phi^{-2}\td S=\frac{A_{\mathrm{ph,out}}}3\,.
\end{equation}
To prove Eq.~\eqref{maxAbs1}, our tool is variational method combining with Lagrangian multipliers. The target functional now is
$$A_S=\int_S\td S\,.$$
According to the conclusion in appendix~\ref{extrms1}, this can be converted into the task  of finding critical surface among the surfaces laying on null sheet $u=0$. Similar to \eqref{maxbs1}, we have following relationship
\begin{equation}\label{maxbs2}
  (\max A_S\mathrm{~on~}\Sigma_t)\leq(\mathrm{maximal~extreme~value~of~}A_S\mathrm{~on~null~sheet~}u=0)\nonumber\,.
\end{equation}
As we want to find the upper bound of $A_S$ when $S$ is a photon sphere, there are a few of restrictions on surface $S$.

Firstly, in the equal-$t$ surface, we have know that a photon sphere is a critical surface which makes the integration $\int\phi^{-2}\td S$ to be extremal. In the null sheet $u=0$, appendix~\ref{extrms1} shows that a photon sphere should also be a critical surface which makes the integration $\int\phi^{-2}\td S$ to be extremal on the null sheet $u=0$. On the null sheet $u=0$, we have $\int\phi^{-2}\td S=\int r_S^3e^{-4\beta}/(Ve^{-2\beta})\td\Omega$. The extreme condition means that $S$ on the null sheet $u=0$ should satisfy following equation
\begin{equation}\label{const1}
  3r_S^{-1}Ve^{-2\beta}-4Ve^{-2\beta}\partial_r\beta-\partial_r(Ve^{-2\beta})=0\,.
\end{equation}
As we now fix the value of functional $\mathcal{B}=\mathcal{B}_0$, we have following constraint
\begin{equation}\label{defLS2}
  \mathcal{B}=A_{S}^{-2}\int_S r^{-1}Ve^{2\beta}\td S=\mathcal{B}_0\,.
\end{equation}
which shows
\begin{equation}\label{defLS2}
  \int r_SVe^{2\beta}\td\Omega-\mathcal{B}_0A_S^2=0\,.
\end{equation}
Thus, we construct following target functional on the null sheet $u=0$ with two Lagrangian multipliers
\begin{eqnarray}\label{defFeq}
F&=&A_S+\int[3r_S^{-1}Ve^{-2\beta}-4Ve^{-2\beta}\partial_r\beta-\partial_r(Ve^{-2\beta})]\lambda_1(x^A)\td\Omega\\
  &&+\lambda_2\left(\int r_SVe^{2\beta}\td\Omega-\mathcal{B}_0A_S^2\right)\nonumber\,.
\end{eqnarray}
Here $\{\lambda_1(x^A)$ $\lambda_2\}$ are two Lagrangian multipliers and $\lambda_2$ is constant. The critical values of $A_{\mathrm{ph,out}}$ in equal-$t$ surface are given by the critical values of $F$ in the null sheet $u=0$.

To solve the variational problem, we separate functional $F$ into two parts
$$F=F_{\mathrm{bd}}+F_{\mathrm{bulk}}\,,$$
where
\begin{equation}\label{defFbd}
  F_{\mathrm{bd}}:=A_S-\lambda_2\mathcal{B}_0A_S^2-\int_S[\partial_r(Ve^{-2\beta})+4(\partial_r\beta)Ve^{-2\beta}]\lambda_1(x^A)\td\Omega
\end{equation}
and
\begin{equation}\label{defFbulk}
  F_{\mathrm{bulk}}:=\int_S\td\Omega[3r_S^{-1}\lambda_1(x^A)+e^{4\beta}\lambda_2]Ve^{-2\beta}\,.
\end{equation}
The $\partial_r(Ve^{-2\beta})$ and $\partial_r\beta$ in $F_{\mathrm{bd}}$ are given by Eq.~\eqref{eqforbeta}. The $Ve^{-2\beta}$ in $F_{\mathrm{nd}}$ is given by Eq.~\eqref{valueV1}. Using Eqs.~\eqref{eqforbeta}, \eqref{eqforV} and \eqref{valueV1} we find
\begin{equation}\label{defrVs1}
  \partial_r(Ve^{-2\beta})|_{r=r_S}=\R/2-\hpD^2\beta_S-(\hpD\beta_S)^2-N_{1S}^2-N_{2S}^2
\end{equation}
and
\begin{equation}\label{eqbetavs1}
  \left.(\partial_r\beta)Ve^{-2\beta}\right|_S=8N_{1S}^2+W_S^2,~~W_S^2:=\left.2\pi r T_{rr}Ve^{-2\beta}\right|_S\,.
\end{equation}
Here $\beta_S=\beta|_S$, $N_{1S}^2$ and $N_{2S}^2$ are defined by Eq.~\eqref{defineM1} but restricted on surface $S$. We use the lower index ``$S$'' to explicitly show that they are surface quantities. Then using the conformal transformation~\eqref{cofrg1}, we have
\begin{eqnarray}\label{fbdeq2}
  F_{\mathrm{bd}}&&=A_S-\lambda_2\mathcal{B}_0A_S^2-\int\lambda_1(y)[1-\hpD^2\Psi_S-\hpD^2\beta_S-(\hpD\beta_S)^2\\
  &&-33e^{-2\Psi}N_{1S}^2-e^{-2\Psi}N_{2S}^2-4e^{-2\Psi}W_S^2]\sqrt{\gamma}\td^2y\nonumber\,.
\end{eqnarray}
and Eq.~\eqref{valueV1} shows
\begin{eqnarray}\label{defFbul2}
  F_{\mathrm{bulk}}&=&\int\td\Omega[3r_S^{-1}\lambda_1+e^{4\beta_S}\lambda_2]Ve^{-2\beta}\\
  &=&\int_{r_{h}}^{\infty}\td r\int\Theta(r_S-r)[3r_S^{-1}\lambda_1+e^{4\beta_S}\lambda_2]\nonumber\\
  &&\times[1-\hpD^2\Psi-\hpD^2\beta-(\hpD\beta)^2-e^{-2\Psi}N_1^2-e^{-2\Psi}N_2^2]\sqrt{\gamma}\td^2y\nonumber\,.
\end{eqnarray}
%
Functional $F_{\mathrm{bd}}$ only involves quantities at the surface $S$, but functional $F_{\mathrm{bulk}}$ involves quantities at surface $S$ as well as in bulk region between $S$ and $\mathcal{H}$. The functional $F$ now depends on 6 surface variables $\{N_{1S},N_{2S},\beta_S, \Psi_S, W_S, r_S\}$ and 4 bulk variables $\{N_1, N_2, \beta, \Psi\}$. We can treat all these variables as independent variables.

The $\Psi_S, W_S$, $N_{1S}$ and $N_{2S}$ appear only in boundary part $F_{\mathrm{bd}}$. Using Eq.~\eqref{fbdeq2}, we find that the variations with respective to $W_S, N_{1S}$ and $N_{2S}$ show
$$N_{1S}=N_{2S}=W_S=0\,.$$
Then variation of $\Psi_S$ reads
\begin{equation}\label{eqforlambda1}
  \hpD^2\lambda_1(y^A)=0\,.
\end{equation}
This shows that $\lambda_1(y^A)$ is a constant. The variation with respective to $\beta_S$ involves both $F_{\mathrm{bulk}}$ and $F_{\mathrm{bd}}$. The result reads,
\begin{equation}\label{eqforbetas2}
  \lambda_1\hpD^2\beta_S=-4\lambda_2e^{4\beta_S}(Ve^{-2\beta})_{r=r_S}\,.
\end{equation}
As $\lambda_1$ is constant, we integrate Eq.~\eqref{eqforbetas2} on $S$ and find
\begin{equation}\label{eqforbetas3}
  0=-4\lambda_2\int e^{4\beta_S}(Ve^{-2\beta})_{r=r_S}\sqrt{\gamma}\td^2y\,.
\end{equation}
As $V>0$, we find $\lambda_2=0$, which implies $\hpD^2\beta_S=0$~\footnote{The other solution is $\lambda_1=0$, which should be abandoned as it leads to the variation with respective to $r_S$ has no solution.} and so $\beta_S$ is constant.

The variables $\{N_1,N_2\}$ appear only in bulk part $F_{\mathrm{bulk}}$. The variation with respective to $N_1$ and $N_2$ shows $N_1=N_2=0$. Now we have
\begin{eqnarray}\label{defFbul2b}
  F_{\mathrm{bulk}}=\lambda_1\int_{r_{h}}^{\infty}\td r\int3\Theta(r_S-r)r_S^{-1}[1-\hpD^2\Psi-\hpD^2\beta-(\hpD\beta)^2]\sqrt{\gamma}\td^2y\,.
\end{eqnarray}
Then we variation with respective to $\Psi$ shows $\hpD^2[\Theta(r_S-r)r_S^{-1}]=0$ and so $r_S$ is constant. The variation of $\beta$ shows $\hpD^2\beta=0$ and so $\beta=\beta(r)$ when $r<r_h<r_S$. Thus, we have following results at the surface $S$
\begin{eqnarray}\label{onshelleq1}
  &&Ve^{-2\beta}|_{\mathrm{on-shell}}=r_S-r_h,~~r_SVe^{2\beta}|_{\mathrm{on-shell}}=r_Se^{4\beta_S}(r_S-r_h),\\
  &&\partial_r(Ve^{-2\beta})|_{\mathrm{on-shell}}=\frac12\R|_S,~~\Psi_S=0\,.\nonumber
\end{eqnarray}
Note that $\Psi_S=0\Rightarrow \R|_S=2$. Take these into Eqs.~\eqref{const1} and \eqref{defLS2} we have
\begin{equation}\label{onshellrs}
  r_h=\frac{2r_S}{3},~~4\pi r_S^2=\frac{e^{4\beta_S}}{3\mathcal{B}_0}
\end{equation}
Thus, we find that $F|_{\mathrm{on-shell}}=e^{4\beta_S}/(3\mathcal{B}_0)$. The on-shell values of target functional are not unique. We then have
\begin{equation}\label{maxFeq1}
  \max F=\max\{e^{4\beta_S}/(3\mathcal{B}_0)|\forall\beta_S\leq0\}=\frac1{3\mathcal{B}_0}\,.
\end{equation}
This proves Eq.~\eqref{maxAbs1} and so we obtain $A_{\mathrm{ph,out}}\leq A_{\mathrm{sh,out}}/3$. It needs to note that above proof is still true for every connected photon spheres, i.e. $A_{\mathrm{ph,out},i}\leq A_{\mathrm{sh,out},i}/3$. Thus, we prove one part of our stronger conjecture~\eqref{goalb2}.

\subsection{Proof of $A_{\mathrm{sh,out}}/3\leq 36\pi M^2$}
The solution of Eq.~\eqref{eqforV} can be written as
\begin{eqnarray}\label{valueV1b2}
  Ve^{-2\beta}|_{r_0}&=&r_0-2M-\int_{r_0}^{\infty}[\R/2-1-\pD^2\beta-(\pD\beta)^2\\
&-&\left.8\pi r^2\rho-\frac{Vre^{-2\beta}}{8}h^{AC}h^{BD}(\partial_rh_{AB})(\partial_rh_{CD})\right]\td r\nonumber\,.
 \end{eqnarray}
This shows
\begin{eqnarray}\label{valueV1b3}
  M&=&\frac{r_0}2-\frac1{8\pi}\int_{r=r_0}\td\Omega\left\{Ve^{-2\beta}-\int_{r_0}^{\infty}[(\pD\beta)^2+8\pi r^2\rho\right.\\
&+&\left.\left.\frac{Vre^{-2\beta}}{8}h^{AC}h^{BD}(\partial_rh_{AB})(\partial_rh_{CD})\right]\td r\right\}\nonumber\,.
 \end{eqnarray}
In the gauge that horizon has constant $r=r_h$, we find
\begin{eqnarray}\label{valueV1b3b}
  M&=&\frac{r_h}2+\frac1{8\pi}\int_{r=r_h}\td\Omega\left\{\int_{r_h}^{\infty}[(\pD\beta)^2+8\pi r^2\rho\right.\\
&+&\left.\left.\frac{Vre^{-2\beta}}{8}h^{AC}h^{BD}(\partial_rh_{AB})(\partial_rh_{CD})\right]\td r\right\}\geq\frac{r_h}2\nonumber\,.
 \end{eqnarray}
This shows the Penrose inequality $9A_{\mathcal{H}}/4\leq 36\pi M^2$.

To show $A_{\mathrm{sh,out}}/3\leq 36\pi M^2$, we immerse the photon sphere into the null sheet $u=0$ and take the gauge that the radius of outmost photon sphere $r_{\mathcal{S}}$ is constant. The size of shadow reads
\begin{equation}\label{maphs1}
  A_{\mathrm{sh,out}}=r^{3}_{\mathcal{S}}\int_{r=r_{\mathcal{S}}}\frac{\td\Omega}{Ve^{2\beta}}\,.
\end{equation}
In general, the horizon radius $r_h$ is no longer constant. Eq.~\eqref{valueV1b3} becomes
\begin{eqnarray}\label{valueV1b3c}
  M&=&\frac{r_{\mathcal{S}}}2-\frac1{8\pi}\int_{r=r_{\mathcal{S}}}\td\Omega\left\{Ve^{-2\beta}-\int_{r_{\mathcal{S}}}^{\infty}[(\pD\beta)^2+8\pi r^2\rho\right.\\
&+&\left.\left.\frac{Vre^{-2\beta}}{8}h^{AC}h^{BD}(\partial_rh_{AB})(\partial_rh_{CD})\right]\td r\right\}\nonumber\,.
\end{eqnarray}
In following we will use variational method to show that, in all black holes of which the size of shadow is $A_{\mathrm{sh,out}}=12\pi a_0^2$, the minimal value of mass is $a_0/3$. The main idea is as follow.

We treat $M$ as a functional, the positive mass theorem shows $M\geq0$. Thus, the functional $M$ is bounded from below. Then the minimum value can be found according to variational problem.  As $\mathcal{S}_{\mathrm{out}}$ is a photon sphere, so constraint~\eqref{const1} should also satisfied. We construct the target functional
\begin{eqnarray}\label{valueV1b4}
  \mathcal{M}&=&\frac{r_{\mathcal{S}}}2-\frac1{8\pi}\int_{r=r_{\mathcal{S}}}\td\Omega\left\{Ve^{-2\beta}-\int_{r_{\mathcal{S}}}^{\infty}[(\pD\beta)^2+8\pi r^2\rho\right.\\
&+&\left.\left.\frac{Vre^{-2\beta}}{8}h^{AC}h^{BD}(\partial_rh_{AB})(\partial_rh_{CD})\right]\td r\right\}+\frac{\lambda_2}{4\pi}\left(r^{3}_{\mathcal{S}}\int_{r=r_{\mathcal{S}}}\frac{\td\Omega}{Ve^{2\beta}}-12\pi a_0^2\right)\nonumber\\
&+&\int_{r=r_{\mathcal{S}}}\left[3r_{\mathcal{S}}^{-1}Ve^{-2\beta}-4Ve^{-2\beta}\partial_r\beta-\partial_r(Ve^{-2\beta})\right]\lambda_1(x^A)\td\Omega\,.\nonumber
\end{eqnarray}
Here $\lambda_1(x^A)$ and $\lambda_2$ are two Lagrangian multipliers and $\lambda_2$ is constant. The $\lambda_2$ insures that the size of shadow is fixed to be $12\pi a_0^2$  and the $\lambda_1$ insures that condition~\eqref{const1} is satisfied.

We use the variable transformations~\eqref{defineM1}
and find
\begin{eqnarray}\label{valueV1b4}
  \mathcal{M}&=&\frac{r_{\mathcal{S}}}2-\frac1{8\pi}\int\td\Omega Ve^{-2\beta}+\frac{\lambda_2}{4\pi}\left(r^{3}_{\mathcal{S}}\int_{r=r_{\mathcal{S}}}\frac{\td\Omega}{Ve^{2\beta}}-12\pi a_0^2\right)\\
  &+&\int_{r=r_{\mathcal{S}}}\left[3r_{\mathcal{S}}^{-1}Ve^{-2\beta}-4Ve^{-2\beta}\partial_r\beta-\partial_r(Ve^{-2\beta})\right]\lambda_1(x^A)\td\Omega\nonumber\\
&&+\frac1{8\pi}\int_{r_{\mathcal{S}}}^{\infty}\td r\int\td\Omega[(\pD\beta)^2+N_1^2+N_2^1]\,\nonumber
\end{eqnarray}
The functional $\mathcal{M}$ contains the boundary part and bulk part. In principle, we can follow the similar step of subsection~\ref{subseca2} to solve the variational problem. However, in this case, we  have simpler method.

We first choose $\beta, N_1, N_2$ and $K:=\sqrt{Ve^{-2\beta}}$ as four independent bulk variables. The bulk variations with respective to $\beta, N_1$ and $N_2$ only involves the third line of Eq.~\eqref{valueV1b4}. The result shows
$$r\in[r_{\mathcal{S}},\infty),~~\beta=\beta(r),~~N_1=N_2=0\Rightarrow\partial_rh_{AB}=0\,.$$
Thus we see $h_{AB}=h_{AB}|_{r=\infty}$. This shows $\R=\R|_{r\rightarrow\infty}=2$. Then we find
$$Ve^{-2\beta}=r-2M\,.$$
Thus, the on-shell geometry outside photon sphere $\mathcal{S}_{\mathrm{out}}$ becomes spherically symmetric. We then find that
\begin{equation}\label{maxonshellm1}
  \min M=\min\mathcal{M}|_{\mathrm{on-shell}}=\min\mathcal{M}|_{\mathrm{spherically~symmetric}}\,.
\end{equation}
In spherically symmetric case, we have shown that $\min M=\sqrt{A_{\mathrm{ph,out}}/(108\pi)}=a_0/3$. Thus Eq.~\eqref{maxonshellm1} shows $\min M=a_0/3$. Then we finish the proof of $A_{\mathrm{sh,out}}/3\leq 36\pi M^2$ in the case without spherical symmetry. Note that above proof also implies that $A_{\mathrm{sh,out}}/3=36\pi M^2$ is true only if the geometry outside $\mathcal{S}_{\mathrm{out}}$ is Schwarzschild.

Now we make a short summary on this section. We use variational method to prove $9A_{\mathcal{H}}/4\leq A_{\mathrm{sh,out}}/3$ and $A_{\mathrm{ph,out}}\leq A_{\mathrm{sh,out}}/3\leq36\pi M^2$ under two assumptions: (1) weak and strong energy condition are satisfied, and (2) outmost horizon and outmost photon sphere are connected and smooth. Our proof also implies a rigidity theorem: $A_{\mathrm{ph,out}}=36\pi M^2$ only if the geometry outside outmost photon sphere  is Schwarzschild. Our proof on $A_{\mathrm{ph,out}}\leq A_{\mathrm{sh,out}}/3$ is also true when outmost photon sphere is not connected. As the result, we obtain $A_{\mathrm{ph,out},i}\leq A_{\mathrm{sh,out},i}/3$ for every connected branch and so prove a part of our stronger conjecture~\eqref{goalb2}. The lower bound $9A_{\mathcal{H}}/4\leq A_{\mathrm{ph,out}}$ in spherically symmetric case involves some non-typical energy condition~\eqref{addpr1}. It is not clear how to generalize it into the case without spherical symmetry. We leave the study of this inequality for the future.

~\\
\textbf{Comment on smoothness: } In above proofs, we assumed that the outmost photon sphere $\mathcal{S}_{\mathrm{out}}$ is smooth. This requirement can be be replaced by piecewise smoothness. The reason is that, in variational method, the target functionals and constraints are defined in integrations. For a outmost photon sphere $\mathcal{S}_{\mathrm{ph,out}}^{(i)}$, we can always find a smooth surface $\mathcal{S}_{\mathrm{out},\varepsilon}^{(i)}$ which satisfies: (1) $\forall\varepsilon>0$, the maximal volume enclosed by $\mathcal{S}_{\mathrm{out}}^{(i)}$ and $\mathcal{S}_{\mathrm{out},\varepsilon}^{(i)}$ is smaller than $\varepsilon M^3$, i.e.
\begin{equation}\label{maxV1}
  \max\int_{V}\td V<\varepsilon M^3\,,
\end{equation}
where $V$ is arbitrary spacelike 3-dimensional surface and satisfies $\partial V=\mathcal{S}_{\mathrm{out}}^{(i)}\cup\mathcal{S}_{\mathrm{out},\varepsilon}^{(i)}$,
and (2) $A_{\mathrm{ph,out},i}$ and $A_{\mathrm{sh,out},i}$ are approximated in arbitrary accuracy
\begin{equation}\label{appros1}
\left|A_{\mathrm{ph,out},i}-\int_{\mathcal{S}_{\mathrm{out},\varepsilon}^{(i)}}\td S\right|<\varepsilon M^2, ~~\left|A_{\mathrm{sh,out},i}-\int_{\mathcal{S}_{\mathrm{out},\varepsilon}^{(i)}}\phi^{-2}\td S\right|<\varepsilon M^2\,,
\end{equation}
and (3) the constraint~\eqref{const1} is broken arbitrarily small
\begin{equation}\label{appros2}
\int_{\mathcal{S}_{\mathrm{out},\varepsilon}^{(i)}}\left|3r^{-1}Ve^{-2\beta}-4Ve^{-2\beta}\partial_r\beta-\partial_r(Ve^{-2\beta})\right|\td S<\varepsilon M^2\,.
\end{equation}
We can use the smooth surface $\mathcal{S}_{\mathrm{out},\varepsilon}^{(i)}$ to replace $\mathcal{S}_{\mathrm{out}}^{(i)}$ and obtain the same conclusion. Because of this reason, the conclusions in our above proofs are still valid if assume the outmost photon sphere is piecewise smooth.

\section{Conclusion}

In this paper, we conjectured a series of universal inequalities about the size of a static black hole in Einstein gravity. We gave a complete proof in the spherically symmetric case. We studied the properties of the photon spheres in general static spacetimes and proved that photon spheres are conformal invariant structures of the spacetime. Our results strongly suggest that black holes photon spheres may have rich physical contents and mathematical structures.

Our conjecture gives us a simple way to estimate the size of the horizon and black hole mass. For the spherically symmetric case, though we assume the spacetime is static outside the horizon (if exists), we in fact only need it being static outside the photon sphere due to Birkhoff theorem. It needs to emphasize that the upper bound in \eqref{unverineq} do not require the existence of a black hole. This has significance in astronomy. Birkhoff theorem implies the interior of photon sphere may not contain a black hole. For example, a neutron star can also form a photon sphere and the corresponding shadow. However, our inequality~\eqref{unverineq} implies if the radius of photon sphere is larger than $2.25M_{\odot}$ or the radius of shadow is larger than $3.89M_{\odot}$, then the interior of the photon sphere cannot be a neutron star. If we find a larger size photon sphere or shadow, the interior must be a black hole or it is to form a black hole. In astrophysical observations, to verify whether or not a gravitational system has photon sphere can give us useful information about the gravitational system. For example, if the spacetime contains a naked singularity rather than a black hole, we can classify naked singularities into two kinds based on whether or not a naked singularity is covered within a photon sphere~\cite{Virbhadra:2002ju}.  For the case which contains a photon sphere, the gravitational lensing effects of naked singularity is very similar to Schwarzschild black hole; however, for the case without photon sphere, gravitational lensing effects is qualitatively different from the case of a Schwarzschild black hole. The combinational observations between the shadow and other gravitational lensing effects in principle can gives us some physical information of gravitational source such as the charge/mass ration, distance to the gravitational source and so on~\cite{Virbhadra:1998dy,Virbhadra:2007kw,Virbhadra:2008ws}. Thus, we can study how to read these useful physical information from the observations on shadows of black holes in the future.

\acknowledgments
We are grateful to Zhong-Ying Fan and Jun-Bao Wu for useful discussions. This work is supported in part by NSFC (National Natural Science Foundation of China) Grant No.~11875200 and No.~11935009.

\appendix
\section{Static gauge in Bondi-Scahs formalism}\label{staticBS}
Consider the metric in Bondi-Scahs formalism
\begin{equation}\label{BSmetricsq1}
  \td s^2=-(\phi^2+|U|^2)\td u^2-2e^{2\beta}\td u\td r+r^2h_{AB}(\td x^A-U^A\td u)(\td x^B-U^B\td u)\,.
\end{equation}
where $(\partial/\partial u)^\mu=\xi^\mu$ is the static Killing vector and $|U|:=r^2h_{AB}U^AU^B$. For the same static spacetime, we can also choose the infalling null time $v$, which  satisfies $(\partial/\partial v)^\mu=\xi^\mu$ and keeps $\{r, x^A\}$ coordinates unchanged. As the result, the metric becomes
\begin{equation}\label{BSmetricsq2}
  \td s^2=-(\phi^2+|\tilde{U}|^2)\td v^2+2e^{2\tilde{\beta}}\td v\td r+r^2\tilde{h}_{AB}(\td x^A+\tilde{U}^A\td v)(\td x^B+\tilde{U}^B\td v)\,.
\end{equation}
Here all components are independent of $v$ and $|\tilde{U}|:=r^2\tilde{h}_{AB}\tilde{U}^A\tilde{U}^B$. As vectors $(\td x^A)_\mu$ and $(\td r)_\nu$ are unchanged in two different coordinates, we have~\footnote{In general, the vectors $(\partial/\partial r)^\mu$ and $(\partial/\partial x^A)^\mu$ in odd coordinates will be different from the vectors $(\partial/\partial r)^\mu$ and $(\partial/\partial x^A)^\mu$ in new coordinates. }
\begin{equation}\label{eqU1}
  g^{\mu\nu}(\td x^A)_\mu(\td r)_\nu=-U^Ae^{-2\beta}=\tilde{U}^Ae^{-2\tilde{\beta}}
\end{equation}
and
\begin{equation}\label{eqforhr}
  g^{\mu\nu}(\td x^A)_\mu(\td x^B)_\nu=r^{-2}h^{AB}=r^2\tilde{h}^{AB}\Rightarrow h_{AB}=\tilde{h}_{AB}\,,
\end{equation}
Define $\tilde{v}=-v$ and we obtain
\begin{equation}\label{BSmetricsq2}
  \td s^2=-(\phi^2+|\tilde{U}|^2)\td \tilde{v}^2-2e^{2\tilde{\beta}}\td \tilde{v}\td r+r^2h_{AB}(\td x^A-\tilde{U}^A\td \tilde{v})(\td x^B-\tilde{U}^B\td \tilde{v})\,.
\end{equation}
This metric and metric~\eqref{BSmetricsq1} describe same geometry at every point only if $\tilde{U}^A=U^A$ and $\beta=\tilde{\beta}$. Combine it with Eq.~\eqref{eqU1} and we find $U^A=0$.

\section{Boundedness of functional $\mathcal{B}$}\label{uppbd}
There is a simple way to prove that $\mathcal{B}$ is bounded from above. Firstly, the area of any closed surface outside horizon is given by
\begin{equation}\label{areaS1}
  A_S=\int_S\td S=\int r_S^2\td\Omega\geq\int r_h^2\td\Omega=4\pi r_h^2=A_{\mathcal{H}}\,.
\end{equation}
If we can prove $0<\phi<1$, we can find $\mathcal{B}<A_S^{-2}\int_S \td S=A_S^{-1}\leq A_{\mathcal{H}}^{-1}$. Thus, in all black holes of fixed horizon area, the functional $\mathcal{B}$ is bounded from above.

In order to prove $0<\phi<1$, we take a 3+1 decomposition
\begin{equation}\label{decomsigam1}
  \td s^2=-\phi^2\td t^2+h_{ab}\td x^a\td x^b\,.
\end{equation}
Here $h_{ab}$ is the metric of equal-$t$ slice $\Sigma_t$. We write down the decomposition of static Einstein's equation in the coordinates gauge~\eqref{decomsigam1}. The Hamiltonian constraint reads
\begin{equation}\label{hamilt}
  R=16\pi \rho\,.
\end{equation}
The momentum constraint and evolutional equation of induced metric are trivial due to extrinsic curvature of $\Sigma_t$ is identical zero. The evolutional equation of extrinsic curvature then shows
\begin{equation}\label{eqdotK1}
  R_{ab}=\phi^{-1}D_aD_b\phi+8\pi[{\mathcal{T}}_{ab}+\frac12h_{ab}(\rho-\mathcal{T})]
\end{equation}
There $R_{ab}$ and $D_a$ are the Ricci tensor and covariant derivative operator of $h_{ab}$, $\mathcal{T}_{ab}$ is the projection of stress tensor at time slice $\Sigma_t$ and $\mathcal{T}$ is the trace of $\mathcal{T}_{ab}$. Eqs.~\eqref{eqdotK1} and \eqref{hamilt} give us the complete description on static spacetime in Einstein gravity. The Combination of Eqs.~\eqref{eqdotK1} and \eqref{hamilt} shows
\begin{equation}\label{eqforphi2}
  D^2\phi=4\pi\phi(\rho+\mathcal{T})\,.
\end{equation}
The strong energy condition implies $\rho+\mathcal{T}\geq0$. Thus, outside horizon $\mathcal{H}$ we have
\begin{equation}\label{eqforphi3}
  D^2\phi\geq0\,.
\end{equation}
Then the maximum principle shows that the maximum of $\phi$ must be at the boundaries, i.e. the spital infinity or outmost horizon $\mathcal{H}$. This show $\phi<1$.

\section{Critical surfaces on equal-$t$ submanifold and $u=0$ null sheet}\label{extrms1}
In this appendix, we will explain why the task of finding critical surface of a target functional on equal-$t$ plane can be converted into the task of finding critical surface of the same target functional on the null sheet $u=0$.

We first consider a concrete example. Assume that $\Sigma_t$ is equal-$t$ submanifold and define a target functional
\begin{equation}\label{defpeq1}
  P:=\int_{S}\phi^{-2}\td S
\end{equation}
Assume that $S_0$ is a critical surface (i.e. the surface makes $P$ extreme) among all surfaces laying on $\Sigma_t$. In following, we will show that, \textit{for arbitrary 3-dimensional manifold $\Gamma$,  if $S_0\subset\Gamma$, then $S_0$ also makes integration $\int_{S}\phi^{-2}\td S$ to be extreme among all surfaces laying on $\Gamma$}.\footnote{This is nontrivial because the critical point of a function(or functional) may be no longer a critical point if we change definition domain of arguments. For example, consider a function $f(x,y,z)=y^2x^2+z$. the point $p_0=\{x=0,y=0,z=0\}$ is a critical point if we restrict us in the $z=0$ plane. However, if we consider an other plane, saying the plane $z=x$, the point $p_0$ is no longer a critical point. }

The proof is as follow. Outside horizon, we can take the orbit of static Killing vector  as the time  and so metric has a form $\td s^2=-\phi^2\td t^2+\td s_{\Sigma_t}^2$. We can take a special gauge for the induced metric in $\Sigma_t$ such that $\td s_{\Sigma_t}^2=\td r^2+q_{AB}\td x^A\td x^B$. This can be achieved locally for arbitrary static black hole. Then the spacetime metric has following 3+1 decomposition form
\begin{equation}\label{spe31ds1}
  \td s^2=-\phi^2\td t^2+\td r^2+q_{AB}\td x^A\td x^B\,.
\end{equation}
A general 3-dimensional submanifold $\Gamma$ is locally parameterized by $t=t_{\Gamma}(r,x^A)$. A surface $S\subset\Gamma$ then is parameterized by $r=r_S(x^A)$ on $\Gamma$. As the result, $S$ is determined by following two functions:
$$r=r_S(x^A),~~t=t_{\Gamma}(r_S(x^A),x^A):=t_S(x^A)\,,$$
and the induced metric of $S$ reads
$$\td s^2=(q_{AB}+\partial_Ar\partial_Br-\phi^2\partial_At_S\partial_Bt_S)\td x^A\td x^B=(\tilde{q}_{AB}-\phi^2\partial_At_S\partial_Bt_S)\td x^A\td x^B\,.$$
Here $\tilde{q}_{AB}:=q_{AB}+\partial_Ar_S\partial_Br_S$, which is independent of $t_S$. It needs to note that $t_S$ depends on the function $r_S(x^A)$ and so
$$\partial_At_S=\partial_rt_{\Gamma}\partial_Ar_S+\partial_At_\Gamma,~~\frac{\partial t_S}{\partial r_S}=\partial_rt_\Gamma\,.$$
Then the target function~\eqref{defpeq1} becomes
\begin{equation}\label{defpeq3}
  P=\int\mathcal{L}\td^2x,~~\mathcal{L}:=\left.\phi^{-2}\sqrt{1-\phi^2\tilde{q}^{AB}\partial_At_S\partial_Bt_S}\sqrt{\tilde{q}}\right|_{r=r_S(x^A)}\,.
\end{equation}
$\tilde{q}^{AB}$ and $\tilde{q}$ are the inverse and determinate of $\tilde{q}_{AB}$, respectively.  If $S$ is the critical surface on $\Gamma$, then we have following equation
\begin{equation}\label{critialeq10}
  \frac{\delta}{\delta r_S}P=\left(\frac{\delta}{\delta r_S}P\right)_{\mathrm{fix}~t_S}+\left(\frac{\delta}{\delta t_S}P\right)_{\mathrm{fix}~r_S}\frac{\partial t_S}{\partial r_S}=0\,.
\end{equation}
Direct computation shows that
$$\left(\frac{\delta}{\delta t_S}P\right)_{\mathrm{fix}~r_S}=\partial_B\left(\frac{\sqrt{\tilde{q}}\tilde{q}^{AB}\partial_Bt_S}{\sqrt{1-\phi^2\tilde{q}^{AB}\partial_At_S\partial_Bt_S}}\right)$$
and so we find Eq.~\eqref{critialeq10} reduces
\begin{equation}\label{critialeq1}
  \frac{\partial\mathcal{L}}{\partial r_S}+\partial_B\left(\frac{\sqrt{\tilde{q}}\tilde{q}^{AB}\partial_Bt_S}{\sqrt{1-\phi^2\tilde{q}^{AB}\partial_At_S\partial_Bt_S}}\right)\partial_rt_\Gamma=0\,.
\end{equation}
For a surface $S$ laying on $\Sigma_t$, its induced metric reads $\td s^2=\tilde{q}_{AB}\td x^A\td x^B$ and so
\begin{equation}\label{defpeq1a}
  P=\int_{r=r_S(x^A)}\phi^{-2}\sqrt{\tilde{q}}\td^2x=\int\mathcal{L}|_{\partial_At_S=0}\td^2x\,.
\end{equation}
The surface $S_0$,which is parameterized by $t_S=$ constant and $r_S=r_0(x^A)$, is critical surface among surfaces laying on $\Sigma_t$, so it satisfies extreme equation
\begin{equation}\label{defpeq2}
 \delta P/\delta r_S=0\Rightarrow\left.\frac{\partial\mathcal{L}}{\partial r_S}\right|_{\partial_At_S=0}=0\,.
\end{equation}
This shows that $S_0$ is also one solution of Eq.~\eqref{critialeq1}. Thus, $S_0$ also makes integration $\int_{S}\phi^{-2}\td S$ to be extreme among all surfaces laying on $\Gamma$ if $S_0$ also lays on $\Gamma$. .

In our above discussion, the 3-dimensional submanifold $\Gamma$ can be spacelike, timelike or null. Assume that the null sheet $u=0$ is formed by the outgoing light rays of surface $S_0$. Above conclusion shows that the task of finding critical surface on equal-$t$ surface can be converted into the task of finding critical surface on the null sheet $u=0$.


Above result can be generalized. Consider a general diffeomorphically invariant functional $\mathcal{F}[S,\phi_i]$ in static spacetime, which is defined on a 2-dimensional surface $S$ and depends on some static fields $\phi_i$ (i.e. $\mathcal{L}_{\xi}\phi_i=0$ and $\phi_i$ is invariant under the time reversal transformation). Assume that $\Gamma$ is arbitrary 3-dimensional hypersurface and surface $S_0\subset\Gamma\cap\Sigma_t$. Using the similar step, we can prove following conclusion: if $S_0$ is a critical surface of $\mathcal{F}[S,\phi_i]$ among the surfaces laying on $\Sigma_t$, it is also a critical surface of $\mathcal{F}[S,\phi_i]$ among the surfaces laying on $\Gamma$.

\section{An intuition about finding upper bound}\label{exampf1}
To explain our logics clearly when we use variational method to prove the inequalities of photon spheres, we will use a concrete function as an example. This example, though is much simpler than the cases in the main content, can exhibit all the necessary ideas.

We consider following 4-variables function
\begin{equation}\label{def3fxyz1}
  f:=\frac{x^2-y^2+\cos^3z+2}{w^2+1}
\end{equation}
This is the analogy of functional $\mathcal{B}$. The four variables of $f$ are not independent. They satisfy following constraint
\begin{equation}\label{constxyz}
  x^2-y^2-\sin^2z+1=0\,.
\end{equation}
This plays the role of Einstein's equation.  It is clear that $f$ has no upper bound if 4 variables are all free. However, only 3 of them are independent due to constraint~\eqref{constxyz}. This is the analogy of the analysis at the end of Sec.~\ref{BSF1}, where we argued that target functionals involves 10 bulk variables and 11 surface variables but only 4 of bulk variables and 8 of surface variables are independent. In following we will prove that $f\leq2$.

Firstly, by using constraint~\eqref{constxyz}, we can prove $f=(1+\cos^3z+\sin^2z)/(w^2+1)<3$. Thus, we find that $f$ is bounded from above. This step is the analogy of appendix~\ref{uppbd}, where we use Einstein's equation to prove that $\mathcal{B}$ is bounded from above.

Secondly, we use constraint~\eqref{constxyz} to rewrite $f$ into a form
$$f=\frac{1+\cos^3z+\sin^2z}{w^2+1}$$
and treat $w,z$ as the independent variables. This is the analogy of the steps between Eq.~\eqref{defLS} to Eq.~\eqref{defy1}, where we used Einstein's equation to rewrite the formula of target functional and chose the independent variables suitably.

Thirdly, we make the variation with respective to $w$ and $z$, and find all critical values (including when $w=\infty$ and $z=\infty$). The variation with respective to $w$ shows that
$$w=0,~~\mathrm{or}~w=\infty\,.$$
When $w=\infty$, we find that on-shell value of $f$ is 0. When $w=0$, the variation with respective to $z$ shows that
$$\sin z=0,~~\mathrm{or}~\cos^2z=2/3\,,$$
which shows
$$f|_{\mathrm{on-shell}}=2,~4/3\pm(2/3)^{3/2}=4/3\pm0.54432\cdots\,.$$
Thus, we find that
$$\max f=\max\{0,2,4/3\pm(2/3)^{3/2}\}=2$$
so we prove $f\leq 2$. This is the analogy of steps between Eq.~\eqref{eqforPsi0} and \eqref{onshellL1}.

To conclude, our basic ideas are as follows. By using constrained equations (in our main text, they are Einstein's equation and requirements of photon sphere), we (1) analyze how many independent variables the target function/functional has, (2) prove target function/functional is bounded from above/blow, (3) rewrite the mathematical expression of target function/functional into a suitable form and choose independent variables suitably, and (4) compute all critical values and find the maximal/minimal one.

\bibliographystyle{JHEP}

\end{document}